\documentclass{article}
\usepackage[english]{babel}
\usepackage{amsmath,amssymb,mathrsfs,theorem,subfiles}
\usepackage{graphicx}
\usepackage{tabu}
\usepackage[table]{xcolor}
\usepackage{diagbox}
\usepackage{multirow}
\usepackage{url}
\usepackage{wrapfig}
\usepackage{setspace}

\usepackage[top=30truemm, bottom=30truemm,left=30truemm,right=30truemm]{geometry}
\def\N{\mathbb N}

\title{Bibliometric analysis on mathematics: \\
3 snapshots: 2005, 2010, 2015} 

\begin{document}

\author{Serge Richard\footnote{Supported by the grant\emph{Topological invariants
through scattering theory and noncommutative geometry} from Nagoya University,
and by JSPS Grant-in-Aid for scientific research C no 18K03328, and on
leave of absence from Univ.~Lyon, Universit\'e Claude Bernard Lyon 1, CNRS UMR 5208,
Institut Camille Jordan, 43 blvd.~du 11 novembre 1918, F-69622 Villeurbanne cedex,
France.},
Qiwen Sun}

\date{\small}
\maketitle
\vspace{-1cm}

\begin{quote}
\emph{
\begin{enumerate}
\item[] Graduate school of mathematics, Nagoya University,
Chikusa-ku, Nagoya 464-8602, Japan
\item[]\emph{E-mails:} richard@math.nagoya-u.ac.jp, m17046h@math.nagoya-u.ac.jp
\end{enumerate}
}
\end{quote}















\begin{abstract}
We carry out a thorough bibliometric analysis of recent publications
in mathematics based on the database \emph{Web of Science}.
The individual relations between various features and the 
\emph{citations} are provided, and the importance of the features
is investigated with decision trees.
The evolution of the features over a period of 10 years
is also studied.
National and international collaborations are scrutinized,
but personal information are fully disregarded.
\paragraph{keywords:} Citations, bibliometric predictors, mathematics, tree-based methods, international collaborations.
\end{abstract}

\section{Introduction}

This research has been triggered by a simple question: Do international
collaborations increase the number of citations in mathematics\;\!?
By looking at the existing studies about this question in various
fields of research \cite{LFSEM,SWBA,WWL,WTG,WZJZ}, the easy and naive
answer would be positive.
However, some investigations show that a more precise answer
depends on the field of research, and that additional information 
should be taken into account, see for example \cite{GDD,S1,S2}.
Thus, from the original narrow question, our interest has shifted
to the more general question: What are the important predictors
for the publications in mathematics, if the response is the number of citations\;\!?

Similar bibliometric investigations have already been performed, as for example in \cite{DT}, but mathematics was not considered in this reference, and the analysis
is partially model dependent. Specific to mathematics, let us mention
the early studies \cite{G1,G2} based on MathSciNet, followed by
\cite{Bens} which discusses the different citation indices for mathematics journals,
\cite{BL} which performs a bibliometric analysis on the period 1868--2008,
and \cite{OZ} which focuses on mathematics education.
For very recent investigations, let us also mention
\cite{Szo} which also studies citations but from an individual perspective,
\cite{Verma} which provides a detailed bibliometric analysis over 40 years but on publications of a single journal,
and \cite{PI} which studies US mathematics faculties with some bibliometric tools.

In order to provide a broad picture about recent publications in mathematics,
and therefore complement some of the publications introduced above,
our initial hope was to use MathSciNet, which is familiar to all mathematicians
and which really focus on mathematics publications. Unfortunately,
MathSciNet does not allow any automated searching or downloading,
and collecting enough information for any serious analysis turns out to be impossible
(despite several requests). On the other hand, Web of Science, which is not specific
to mathematics but contains information about mathematics among other fields, allows the collect of large amount of data, and its supporting team answered all our inquiries.
For these reasons, after a comparison of the two databases provided in Section
\ref{sec_back}, we concentrate in the subsequent sections on data provided by Web of Science only. 

Let us now be more specific about the content of this paper.
Our investigations are focusing on publications in mathematics for three years: 2005, 2010, and 2015. This choice allows us to see an evolution in the publication records over a period of ten years, without overwhelming us with too much data.
For each of these three years, we collected between 45'000 and nearly 80'000 items related to mathematics, and for each item we kept the record of 10 features as predictors together with the response, namely the number of citations up to November 2020.
These predictors are introduced and discussed in Section \ref{sec_pred}.

The preliminary analysis consists in looking at the response as a function
of a single predictor.
Let us immediately stress that since the response depends on time
(the number of citations increases as years pass), all investigations
are performed on the three years independently.
Section \ref{sec_ind_pre} contains these results, presented either with graphs or with tables. More precisely, the citations are provided successively as a function of
\begin{enumerate}
\item[(i)] the number of authors,
\item[(ii)] the number of countries associated with the authors,
\item[(iii)] the number of institutes associated with the authors,
\item[(iv)] the number of references provided by the authors,
\item[(v)] the number of pages of the publication,
\item[(vi)] the number of keywords provided by the authors
\item[(vii)] the open access (or not) of the publication,
\item[(viii)] the journal impact factor JIF (if the publication has appeared in a journal with a JIF),
\item[(ix)] the research area of the publication,
\item[(x)] the categories associated with the publication.
\end{enumerate}
More explanations and comments are provided in Section \ref{sec_ind_pre}.

It clearly appears in these individual investigations that the response is related
to some predictors, but how much information can be extracted from them,
and what is their relative importance\;\!?
These questions, and others, are discussed in Section \ref{invest}.
Because of the diversity of the predictors we opted for an approach based on
tree-based methods, as introduced in \cite{BFOS}.
Indeed, unlike the approach provided in \cite{SWBA}
we do not want to consider some linear relations between the predictors and the
response, but prefer an approach which divides the predictor space into several
regions and associates to each region a local response.
Alternatively, we could have borrowed some \emph{bibliometrix} tools developed in
\cite{AC} if our investigations were performed on R, but we opted for the tools
available on the platform \cite{Gini}.

Several experiments are performed with trees, with some parameters chosen according
to the year of publications and to the existence (or non-existence) of a JIF associated with the publications. Based on these experiments, the predictors can be ranked according to their importance. Another outcome of tree classifiers is the ability of predicting the citations (at least within some predefined classes) based on the predictors. Clearly, the result is not very good, but the converse would have been even more surprising. However, the predictions are better than a random guess, as
explained in Section \ref{invest}.

In Section \ref{sec_countries} we turn our attention to countries:
What information can be deduced from the individual publications about the research in the countries of the authors\;\!?
Can one measure a kind of performance for each country\;\!? And what about collaboration between countries, which is related to our very initial question, can one measure these collaborations, and say something about them\;\!? Data for answering these questions are presented in Section \ref{sec_countries} for the main countries, which means for the country producing the majority of publications in mathematics. In fact, data covering about 130 countries were available, but for some of them, the annual number of publications is too limited to support any analysis.

With this paper we provide three snapshots (2005, 2010, 2015) about the publications in mathematics, and
extract as much bibliometric information as possible.
As already mentioned, we would have preferred working on a database MathSciNet because some information would have been more accurate. It is quite unfortunate that the policy and the tools provided by this website do not allow such investigations,
as implicitly acknowledged in \cite{Dunne}.
On the other hand, by using Web of Science database, our investigations about mathematics have covered a slightly broader range of publications.

\section{General pictures}\label{sec_back}

In this section we provide general information about publications in mathematics for the last 20 years. A few comparisons between the two databases
MathSciNet (MSN) and Web of Science (WoS) are also presented.
Finally, the data we shall use in the following sections are introduced, and some statistics are exhibited.

\begin{wrapfigure}{R}{0.5\textwidth}
\centering
\includegraphics[scale = 0.4]{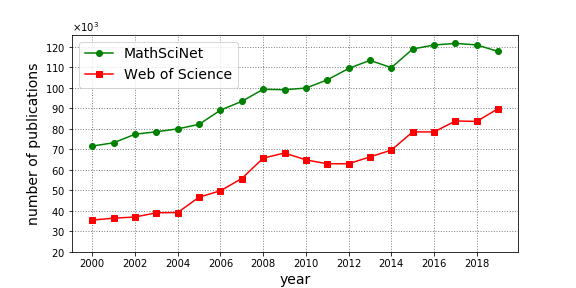}
\caption{Yearly new indexed math publications}
\label{fig:glo_num_pub}
\end{wrapfigure}

Since this research is based on data about mathematics,
let us first have a look at two important sources of information.
MathSciNet is an electronic database operated by the American Mathematical Society
focusing exclusively on publications in mathematics. In November 2020 it contains about 3.9 millions items.
Web of Science is a much more general database operated by the private company Clarivate. It is possible to select publications in mathematics by choosing the \emph{research area} mathematics (SU=mathematics). In November 2020, the outcome for this general request is about 2.0 millions items. Note that WoS contains also \emph{categories}, and one of them corresponds to mathematics. However, by choosing this request (WC=mathematics) the number of items is 1.7 millions, and these items are strictly contained in the previous request about research area.

\begin{figure}[h]
\centering
\begin{minipage}[b]{0.4\textwidth}
\includegraphics[scale = 0.4]{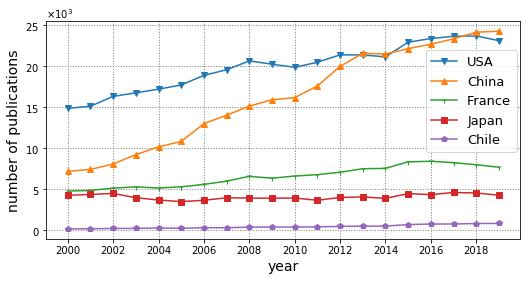}
\end{minipage}%
\hfill
\begin{minipage}[b]{0.5\textwidth}
\includegraphics[scale = 0.4]{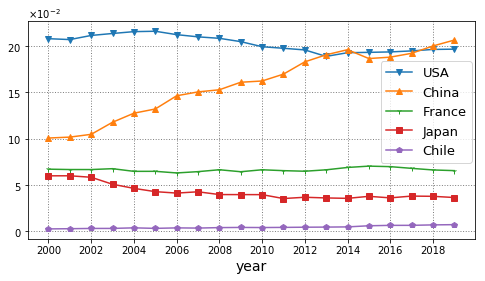}
\end{minipage}
\caption{Publications with at least one author from a given country: absolute and relative numbers from MSN}
\label{fig:con_msn}
\end{figure}

\begin{figure}[h]
\centering
\begin{minipage}[b]{0.4\textwidth}
\includegraphics[scale = 0.4]{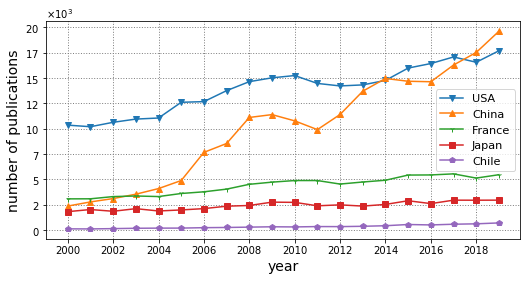}
\end{minipage}%
\hfill
\begin{minipage}[b]{0.4\textwidth}
\includegraphics[scale = 0.4]{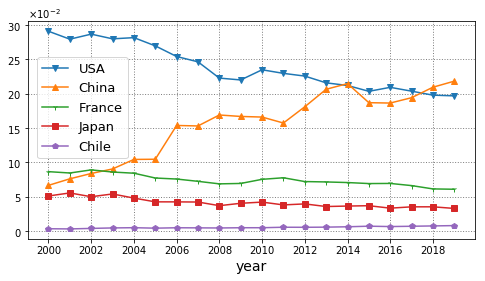}
\end{minipage}
\caption{Publications with at least one author from a given country: absolute and relative numbers from WoS}
\label{fig:con_wos}
\end{figure}

A comparison about the number of works published in the last 20 years is provided
in Figure \ref{fig:glo_num_pub}. For MSN, all works are reported, while for WoS
it is again the works corresponding to the research area mathematics.
Let us provide one more comparison between the two databases, based on one
information that will be used in the analysis.
The information is related to the country in which research institutions (universities, research institutes, etc) are located. For simplicity, we shall call this the \emph{country} of the research institution, and by extension the country of the author working in this research institution.
Figures \ref{fig:con_msn} and \ref{fig:con_wos} show the yearly publications and their relative numbers with at least one author from
a research institution in one of the following countries: USA, China, France, Japan, Chile.
The relative numbers are with respect to the total number of publications in mathematics index by MSN and WoS (shown in Figure \ref{fig:glo_num_pub}).
Thus, even if the difference between the total numbers of items in the two databases is not negligible, we expect that their shapes and trends are similar.

\begin{figure}[h]
\centering
\includegraphics[scale = 0.50]{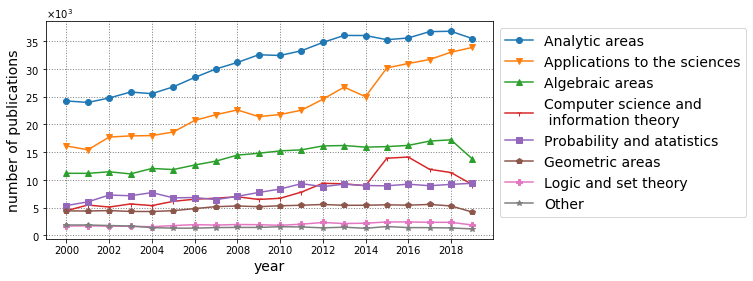}
\caption{Yearly new indexed math publications (8 fields) based on the MSC}
\label{fig:glo_sub}
\end{figure}

On the other hand, a unique feature of MSN is the Mathematics Subject Classification (MSC).
This classification contains more than 60 subjects, and each one can be divided into numerous sub-subjects. Each publication is indexed by one primary subject or one primary subject with several secondary subjects. The subjects are usually chosen by the author(s) of a publication, or carefully assigned by the editors of MSN. The MSC provides a rather precise information about the content of each publication. With this information, a refined plot of Figure \ref{fig:glo_num_pub} for MSN is shown in Figure \ref{fig:glo_sub}.
The eight combined subjects are elaborated in \cite{RD}.
Unfortunately, WoS does not contain the MSC, and therefore we shall not be able to use this information in our investigations.

As mentioned in the Introduction, our analysis is based on the data in three years. With the request \emph{research area = mathematics}, and once the items with no clear author or with no clear affiliation for the author(s) have been removed, the number of items collected are:
\begin{equation}\label{eq_step_2}
\begin{split}
& \hbox{research area = mathematics}, \hbox{clear author(s) and affiliation(s)} \\
& 2005:\ 45'035 \hbox{ items} \qquad
2010:\ 62'945 \hbox{ items} \qquad
2015:\ 76'788 \hbox{ items.}
\end{split}
\end{equation}

\begin{figure}[h]
\centering
\begin{minipage}[b]{0.33\textwidth}
\centering
\includegraphics[scale = 0.4]{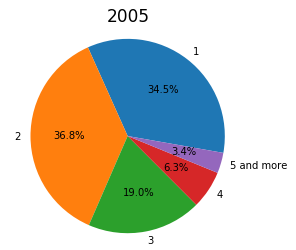}
\end{minipage}%
\begin{minipage}[b]{0.33\textwidth}
\centering
\includegraphics[scale = 0.4]{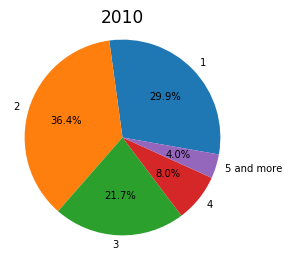}
\end{minipage}%
\begin{minipage}[b]{0.33\textwidth}
\centering
\includegraphics[scale = 0.4]{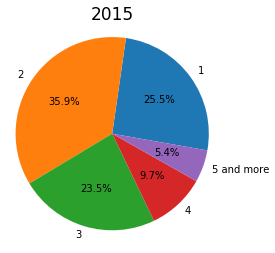}
\end{minipage}
\caption{Distribution of the number of authors per publication}
\label{fig_authors}
\end{figure}

The following statistics are computed on these numerous items.
It has been observed in earlier publications that the average number of authors for each paper has been increasing over time, see for example \cite[Figure 10]{BL}. Since our investigations are based on three distinct years, let us observe this effect on a period of 10 years, see Figure \ref{fig_authors}.

It clearly appears that the proportions of publications with 1 and 2 authors are decreasing, while the ones with 3, 4, or 5 and more authors are increasing. The average number of authors for these three years and based on the data mentioned in \eqref{eq_step_2} are respectively:
\begin{equation}\label{eq_authors}
2005:\ 2.10 \hbox{ authors} \qquad
2010:\ 2.23 \hbox{ authors} \qquad
2015:\ 2.39 \hbox{ authors.}
\end{equation}

Note that other numbers confirm this increase in the collaborations for each publication. Indeed, if one looks at the average number of research institutes involved for each publication one gets
\begin{equation}\label{eq_institutes}
2005:\ 1.57 \hbox{ institutes} \qquad
2010:\ 1.74 \hbox{ institutes} \qquad
2015:\ 1.90 \hbox{ institutes.}
\end{equation}
These numbers have been computed by counting the number of different addresses provided by the publications.

\begin{figure}[h]
\centering
\begin{minipage}[b]{0.33\textwidth}
\centering
\includegraphics[scale = 0.4]{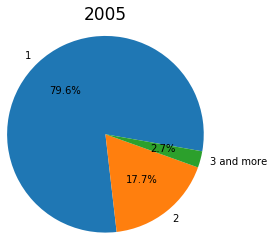}
\end{minipage}%
\begin{minipage}[b]{0.33\textwidth}
\centering
\includegraphics[scale = 0.4]{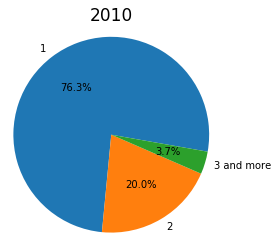}
\end{minipage}%
\begin{minipage}[b]{0.33\textwidth}
\centering
\includegraphics[scale = 0.4]{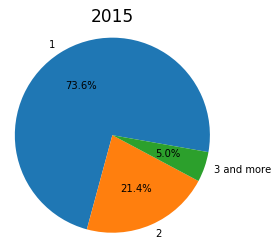}
\end{minipage}
\caption{Distribution of the number of countries per publication}
\label{fig_countries}
\end{figure}

Since one of our interests is to study international collaborations, let us provide similar results for the average number of countries involved for each publication:
\begin{equation}\label{eq_nationalities}
2005:\ 1.23 \hbox{ countries} \qquad
2010:\ 1.28 \hbox{ countries} \qquad
2015:\ 1.32 \hbox{ countries.}
\end{equation}
Again, these numbers have been computed by counting the number of different countries mentioned in the list of addresses of the authors. If we look at the details, one obtains the distributions provided in Figure \ref{fig_countries}.

\section{The predictors and the response}\label{sec_pred}

In this section we introduce the predictors and the response that we have employed
for our investigations, and make a few comments about them.

The predictors can be roughly divided into three categories, namely
those related to the author(s) of a publication, those related to the publication
itself, and those related to the journal or the physical support in which the publication has appeared. All of them have been extracted from the WoS database for the items mentioned in \eqref{eq_step_2}.
Let us immediately stress that the author's names have been completely disregarded
in our investigations.

\begin{itemize}
\item Author(s)
\item[${}$] \emph{authors}: the number of authors
\item[${}$]\emph{countries}: the number of countries
\item[${}$]\emph{institutes}: the number of research institutes
\end{itemize}

Note that in addition to the number of countries involved for each publication, the exact list of countries will also be investigated in Section \ref{sec_countries}.

\begin{itemize}
\item Publication
\item[${}$]\emph{references}: the number of references
\item[${}$]\emph{pages}: the number of pages
\item[${}$]\emph{keywords}: the number of keywords provided by the authors
\item[${}$]\emph{open access}: open access
\end{itemize}

WoS provides some information about various types of open access.
More precisely, publications which are partially or fully open access are identifiable
in the database. This information is obtained in collaboration with \emph{EndNote Click}, formerly called \emph{Kopernio}. Since there exist various levels of open access,
this predictor will be used cautiously.

\begin{itemize}
\item Journal
\item[${}$]\emph{jif}: journal impact factor
\item[${}$]\emph{research areas}: research areas
\item[${}$]\emph{categories}: categories
\end{itemize}

The \emph{journal impact factor} is computed by WoS and assigned to several journals for each year. More information about its computation and its weaknesses can be found
here\footnote{https://en.wikipedia.org/wiki/Impact\_factor}. As already mentioned,
\emph{research areas} and \emph{categories} are classification indices provided by WoS.
Several research areas and several categories have been assigned to each journal in the WoS database, based on several criteria. Compared with the rather precise definitions of the categories, the research areas are less precisely defined (this has been confirmed by the technical support from Clarivate who answered our inquiries).
Moreover, compared with Mathematics Subject Classification (MSC) from MathSciNet, these two predictors are hugely less precise. Not only these indices are not chosen by the authors, but they are common to all publications in one journal, and their assignment is not so clear. Nevertheless, they provide a vague information which deserves to be collected, and which will be further discussed later on.

For the response, the number of citations for each publication has been recorded. These numbers were collected in October/November 2020. It should be emphasized that these
numbers range between 0 and some very large numbers. For example, one work published in 2005 has been cited up to 11'106 times (12'015 and 18'439 times for the most cited works published in 2010 and 2015).

A few other publications are also cited numerous times, which is quite unlikely in the field of mathematics. Since such publications have an enormous impact in the computations of means, we have decided not to consider them. More precisely, we have decided to keep only the publications with a number of citations strictly below 64. This number corresponds to the lower 95th percentile of the collected publications from 2005. For simplicity, we kept this upper limit of 64 also for the data from 2010 and 2015 (which
corresponds respectively to the lower 97th and 99th percentiles). As a consequence, the data with less than 64 citations are
\begin{equation}\label{eq_step_3}
\begin{split}& \hbox{research area = mathematics}, \hbox{clear author(s) and affiliation(s)} , \hbox{citation } <64 \\
& 2005:\ 42'792 \hbox{ items} \qquad
2010:\ 61'084 \hbox{ items} \qquad
2015:\ 76'168 \hbox{ items.}
\end{split}
\end{equation}
In the subsequent computations, and in particular for computations of means, it is these items which are considered.

Let us finally mention that we could have considered the 95th percentiles for the three years, which means that the upper limit for 2010 and 2015 would have been lower than 64. As a result, the means related to these years would have been slightly smaller. However, since we can not compare directly the citations between publications produced respectively in 2005, 2010 and 2015, the simpler choice of keeping the upper bound 64 instead of keeping 95th percentiles does not affect our investigations.

\section{Individual predictors}\label{sec_ind_pre}

In this section we study the relations of the individual predictors with the number of citations. A comparison between the predictors, based on a tree classifier, will be presented in the next section.

\subsection{Number of authors}

\begin{figure}[h]
\centering
\begin{minipage}[b]{0.5\textwidth}
\centering
\includegraphics[scale = 0.4]{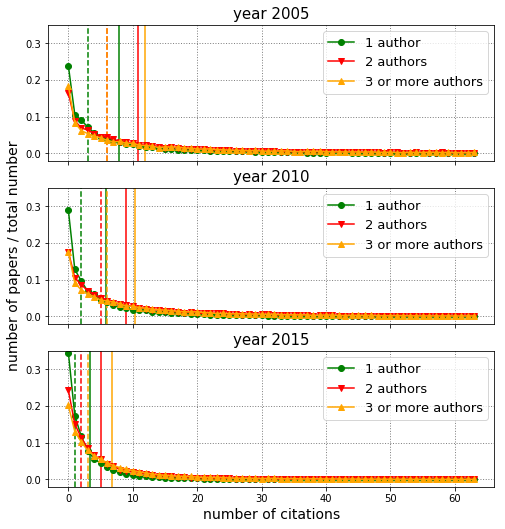}
\end{minipage}%
\begin{minipage}[b]{0.5\textwidth}
\centering
\includegraphics[scale = 0.4]{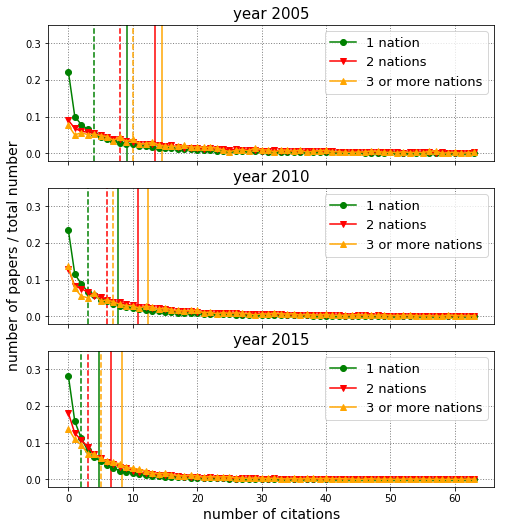}
\end{minipage}%
\caption{Distributions depending on the number of authors or on the number of countries involved}
\label{fig_6_distributions}
\end{figure}

The items mentioned in \eqref{eq_step_3} have been divided according to the number of authors (1, 2, 3 and more) and the respective distributions of citations have been reported in the first column of Figure \ref{fig_6_distributions}.

For each group, the mean has been indicated with a vertical line, and the median has been reported with a vertical dashed line. These precise values are indicated in Table \ref{t_authors_2}, as well as the proportions of publications with 1, 2, or 3 and more authors.

\begin{table}[h]
\centering
\begin{tabular}{|l|c|c|c|c|c|c|c|c|c|}
\hline & \multicolumn{3}{c|}{2005} & \multicolumn{3}{c|}{2010} & \multicolumn{3}{c|}{2015} \\ \hline
& \% & mean & median & \% & mean & median & \% & mean & median \\
\hline 1 author & 35.3 & 7.8 & 3 & 30.4 & 5.9 & 2 & 25.6 & 3.4 & 1 \\
\hline 2 authors & 36.6 & 10.8 & 6 & 36.5 & 9.0 & 5 & 35.9 & 5.1 & 2 \\
\hline $\geq$ 3 authors & 28.1 & 11.8 & 6 & 33.1 & 10.3 & 6 & 38.5 & 6.7 & 3 \\
\hline
\end{tabular}
\caption{Citations depending on the number of authors}
\label{t_authors_2}
\end{table}

On Table \ref{t_authors_2}, it clearly appears that the fractions of publications with 1 or 2 authors decrease over time, while the one with 3 and more authors increases over time. This observation goes in line with the content of Figure \ref{fig_authors}, when no upper limit for the number of citations was imposed. On the other hand, it clearly appears that means and medians of the three years' record increase along with the number of authors.

\subsection{Number of countries}\label{subsection_nb_countries}

Let us now divide the items according to the number of countries appearing in the list of addresses of the authors.
A division into 1, 2, 3 and more countries has also been performed, and the distribution of citations is reported in the second column of Figure \ref{fig_6_distributions}.
The values of the means, the medians, and the proportions are provided in Table \ref{t_countries}.

\begin{table}[h]
\centering
\begin{tabular}{|l|c|c|c|c|c|c|c|c|c|}
\hline & \multicolumn{3}{c|}{2005} & \multicolumn{3}{c|}{2010} & \multicolumn{3}{c|}{2015} \\
\hline & \% & mean & median & \% & mean & median & \% & mean & median \\
\hline 1 country & 80.0 & 9.2 & 4 & 76.5 & 7.7 & 3 & 73.7 & 4.7 & 2 \\
\hline 2 countries & 17.4 & 13.5 & 8 & 19.8 & 10.8 & 6 & 21.4 & 6.6 & 3 \\
\hline $\geq$ 3 countries & 2.6 & 14.5 & 10 & 3.7 & 12.3 & 7 & 4.9 & 8.3 & 5 \\
\hline
\end{tabular}
\caption{Citations depending on the number of countries involoved}
\label{t_countries}
\end{table}

Similar to the content of Table \ref{t_authors_2}, the number of countries involved for each publication increases over time, and the means and medians grow along with the number of countries. The fist observation confirms the trend observed in Figure \ref{fig_countries}. These observations reflect the internationalization of research and publication processes. However, let me mention a small effect inside this global picture. For most items, the appearance of $n$ countries correspond to at least $n$ authors, one (or more) in each country. However, there also exist tens of items with one author having two main addresses in two different countries. Since these situations confirm an additional face of internationalization, we haven't tried to separate these two effects.

\subsection{Number of institutes}

Another division according to the number of different research institutes appearing in the list of addresses is performed.
We divided the items into 1, 2, 3, 4 and more institutes, and computed the proportion, the mean and the median for each of these groups. These numbers are reported in Table \ref{t_institutes}.

\begin{table}[h]
\centering
\begin{tabular}{|l|c|c|c|c|c|c|c|c|c|}
\hline
 & \multicolumn{3}{c|}{2005} & \multicolumn{3}{c|}{2010} & \multicolumn{3}{c|}{2015} \\ \hline
 & \% & mean & median & \% & mean & median & \% & mean & median \\ \hline
1 institute & 60.7 & 8.0 & 3 & 51.7 & 6.8 & 3 & 44.7 & 4.2 & 2 \\ \hline
2 institutes & 27.6 & 12.5 & 8 & 31.0 & 9.6 & 5 & 33.1 & 5.5 & 3 \\ \hline
3 institutes & 8.8 & 14.5 & 9 & 12.4 & 10.8 & 6 & 14.6 & 6.6 & 3 \\ \hline
$\geq$ 4 institutes & 2.9 & 16.5 & 12 & 4.9 & 12.8 & 8 & 7.6 & 8.1 & 4 \\ \hline
\end{tabular}
\caption{Citations depending on the number of institutes}
\label{t_institutes}
\end{table}

As for the previous two predictors, the number of publications involving only one institute decrease over time, while the number of publications involving two and more institutes increase over time. By looking at these three predictors, it appears quite clearly
that an increase in the number of authors, countries, or institutes, corresponds to an
increase in the number of citations. In fact, both the mean and the median are increasing with these predictors.

\subsection{Number of references}

\begin{figure}[h]
\centering
\includegraphics[scale = 0.50]{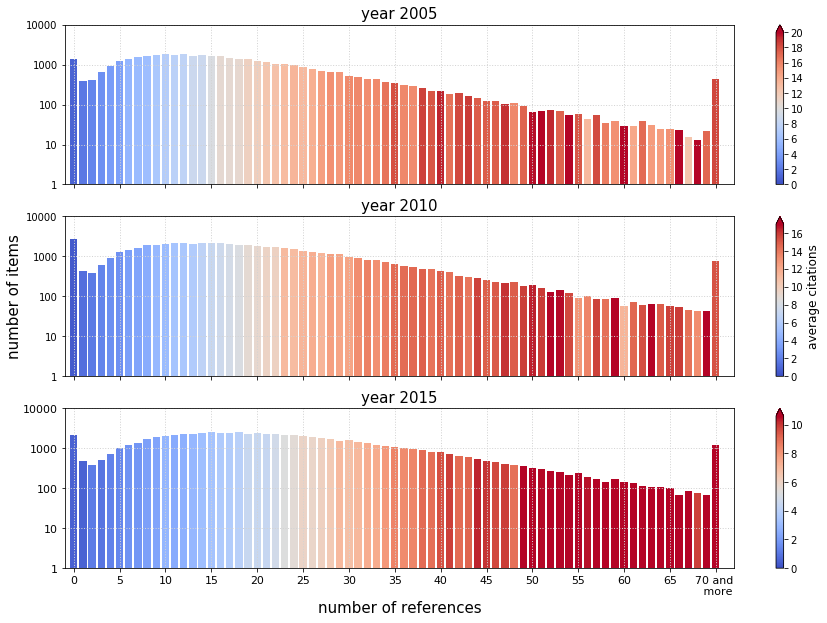}
\caption{Citations depending on the number of references}
\label{fig_references}
\end{figure}

For most items provided by WoS, the number of references mentioned by
the authors of a publication is provided. A classification depending on the number of references has been realized, and Figure \ref{fig_references} provides the information about the number of publications ($y-$axis) with a given number of references ($x-$axis), together with the citation mean (color). Note that in the three graphs, the grey color correspond to the mean citation over all data of the respective year. These means appear in Table \ref{t_ra} (first row), but for the record let us already mention them:
\begin{equation}\label{eq_yearly_mean}
2005:\ 10.0 \hbox{ citations} \qquad
2010:\ 8.5 \hbox{ citations} \qquad
2015:\ 5.3 \hbox{ citations.}
\end{equation}

By computing the average number of references for the 3 years, the expected number of references are:
\begin{equation}\label{eq_references}
2005:\ 18.3 \hbox{ references} \qquad
2010:\ 20.7 \hbox{ references} \qquad
2015:\ 23.8 \hbox{ references.}
\end{equation}

A similar computation can be realized by eliminating the items with $0$ references. Indeed, it is rather doubtful that a publication does not mention any reference, but it seems more likely that WoS has not recorded the information for these publications.
In order to eliminate this bias, we compute the expected number of references, once the items with $0$ references are disregarded, and one gets:
\begin{equation}\label{eq_references_no_0}
2005:\ 19.0 \hbox{ references} \qquad
2010:\ 21.7 \hbox{ references} \qquad
2015:\ 24.5 \hbox{ references.}
\end{equation}

The numbers in \eqref{eq_references} and in \eqref{eq_references_no_0}
show a rather steep increase of the average number of references over a period of 10 years. Also, publications with the mean number of references received about the mean number of citations for each year. It would certainly be
valuable to look at this predictor with data from additional years and over a
longer period of time.

\subsection{Number of pages}

\begin{figure}[h]
\centering
\includegraphics[scale = 0.50]{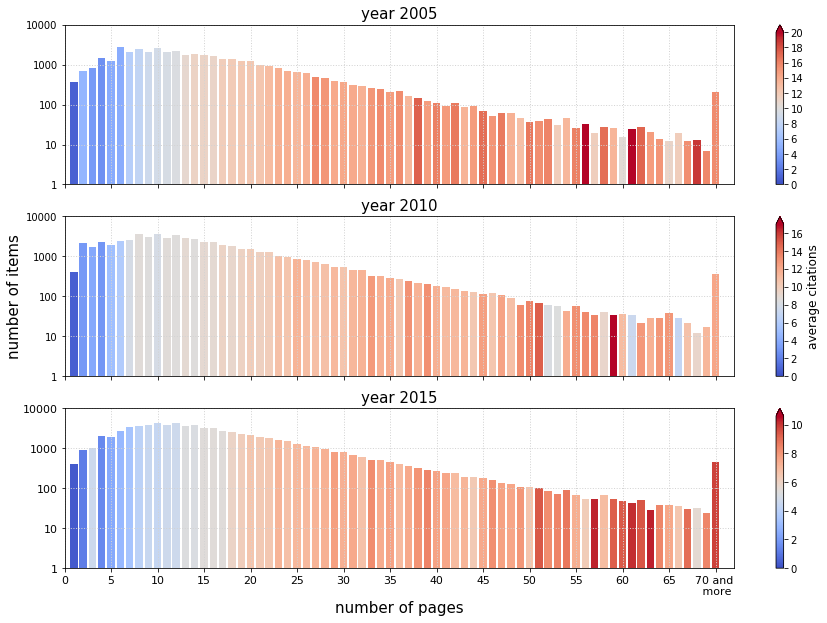}
\caption{Citations depending on the number of pages}
\label{fig_pages}
\end{figure}

Similar investigations can be performed with the number of pages.
Figure \ref{fig_pages} contains the outcomes of this investigation, with the number of items for a given number of pages together with an information about the citation mean.

The conclusion is similar to the one obtained for Figure \ref{fig_references}, namely a positive correlation between the number of pages and the citations.
Once again, the average number of pages can be computed.
The expected numbers of pages are
\begin{equation}\label{eq_pages}
2005:\ 15.5 \hbox{ pages} \qquad
2010:\ 15.5 \hbox{ pages} \qquad
2015:\ 16.9 \hbox{ pages.}
\end{equation}

Compared to the previous result with the number of references, the increasing trend of number of pages is less clear.
Also, one observes that
the contrast of colors is more important in Figure \ref{fig_references} than in Figure \ref{fig_pages}. This difference can be understood as the predominance of the number
of references over the number of pages as a predictor for the number for citations.
This fact will be confirmed in the next section.

\subsection{Number of keywords}

Let us now have a look at keywords. WoS lists the keywords provided by the authors, and a second list of keywords introduced by WoS. Let us immediately say that we were interested only in the first list.
In Table \ref{t_keywords}, we report the proportions of publications as a function
of the number of keywords, and provide also the means and the medians. Note that there is no information about keywords for about $1/3$ of the data for each year. This absence can be due either to the lack of keywords provided in the original support of the publication, or in a failure in the collect of this information by WoS.

\begin{table}[h]
\centering
\begin{tabular}{|l|c|c|c|c|c|c|c|c|c|}
\hline
 & \multicolumn{3}{c|}{2005} & \multicolumn{3}{c|}{2010} & \multicolumn{3}{c|}{2015} \\ \hline
nb keywords & \% & mean & median & \% & mean & median & \% & mean & median \\ \hline
0 & 36.6 & 8.3 & 3 & 28.7 & 6.7 & 2 & 23.7 & 4.4 & 1 \\ \hline
1 & 0.3 & 7.7 & 3 & 0.3 & 4.4 & 2 & 0.2 & 3.1 & 1 \\ \hline
2 & 4.3 & 8.1 & 4 & 3.9 & 6.8 & 3 & 3.0 & 3.4 & 2 \\ \hline
3 & 15.9 & 9.8 & 5 & 16.7 & 7.6 & 4 & 16.2 & 4.3 & 2 \\ \hline
4 & 17.6 & 10.9 & 6 & 20.1 & 8.7 & 4 & 21.9 & 5.4 & 3 \\ \hline
5 & 13.1 & 11.5 & 7 & 15.8 & 10.3 & 6 & 18.6 & 6.2 & 3 \\ \hline
6 & 6.4 & 13.3 & 8 & 7.9 & 11.2 & 6 & 9.5 & 7.0 & 4 \\ \hline
7 & 2.7 & 12.9 & 7 & 3.2 & 10.7 & 6 & 3.3 & 6.3 & 3 \\ \hline
8 & 1.3 & 14.8 & 10 & 1.7 & 11.3 & 7 & 1.7 & 6.2 & 3 \\ \hline
9 & 0.7 & 15.3 & 11 & 0.8 & 11.5 & 7 & 0.9 & 6.9 & 4 \\ \hline
10 & 0.4 & 14.3 & 9 & 0.4 & 12.0 & 6 & 0.5 & 6.0 & 3 \\ \hline
$\geq$ 11 & 0.5 & 15.3 & 11 & 0.5 & 11.2 & 7 & 0.6 & 6.6 & 4 \\ \hline
\end{tabular}
\caption{Citations depending on the number of keywords}
\label{t_keywords}
\end{table}

What appears in this table is a positive relation between the number of keywords
and the citation mean as long as the number of keywords is between 1 and 5.
For the items with more than 5 keywords, any relation between the number of keywords and the citation mean is not really visible.

Let us still provide the average number of keywords. Based on WoS, the expected number of keywords are:
\begin{equation}\label{eq_keywords}
2005:\ 2.78 \hbox{ keywords} \qquad
2010:\ 3.19 \hbox{ keywords} \qquad
2015:\ 3.49 \hbox{ keywords.}
\end{equation}
In order to eliminate the bias related to $0$ keywords,
the statistics with $0$ keywords removed have also been computed,
and one gets
\begin{equation}\label{eq_keywords_no_0}
2005:\ 4.38 \hbox{ keywords} \qquad
2010:\ 4.48 \hbox{ keywords} \qquad
2015:\ 4.58 \hbox{ keywords.}
\end{equation}
As for the number of references and the number of pages, it is visible in
\eqref{eq_keywords} and in \eqref{eq_keywords_no_0} that the number of keywords provided by the authors is also following an increasing trend.

\subsection{Open access}

As already mentioned, WoS lists the items which have a partial or a full open access. Since this information is available, let us just provide the proportion of items having any kind of open access, together with the respective mean and median.
The outcomes are summarized in Table \ref{t_open}.

\begin{table}[h]
\centering
\begin{tabular}{|l|c|c|c|c|c|c|c|c|c|}
\hline
 & \multicolumn{3}{c|}{2005} & \multicolumn{3}{c|}{2010} & \multicolumn{3}{c|}{2015} \\ \hline
 & \% & mean & median & \% & mean & median & \% & mean & median \\ \hline
OA & 16.4 & 12.2 & 7 & 21.2 & 9.7 & 5 & 32.3 & 5.7 & 3 \\ \hline
no OA & 83.6 & 9.6 & 5 & 78.8 & 8.1 & 4 & 67.7 & 5.1 & 2 \\ \hline
\end{tabular}
\caption{Citations with or without open access}
\label{t_open}
\end{table}

The table contains one information which is not surprising: the proportion of items with an open access increases over the years. Also, both the means and the medians are larger for the items with an open access compared to the items without this access. However, even if these results look quite natural, we think that further and more precise information would be necessary to get a better picture about the impact of open access.

\subsection{Journal impact factor}

By definition, the journal impact factor (JIF) is available only for journals, and not for all items. Note that WoS does not link automatically the publications appearing in journals with the corresponding JIF. However, we were able to track this information for about 2/3 of our items. More precisely, a JIF has been associated to the following numbers of items from the list described in \eqref{eq_step_3}:
\begin{equation}\label{eq_step_4}
\begin{split}& \hbox{research area = mathematics}, \hbox{clear author(s) and affiliation(s)} , \hbox{citation } <64, \\
& \hbox{JIF available} \\
& 2005:\ 31'556 \hbox{ items} \qquad
2010:\ 44'639 \hbox{ items} \qquad
2015:\ 57'756 \hbox{ items.}
\end{split}
\end{equation}

It is not surprising that this list of items is biased. Indeed, if we look at the citation means over these items, the numbers are not exactly the ones appearing in
\eqref{eq_yearly_mean} but
\begin{equation}\label{eq_yearly_mean_JIF}
2005:\ 12.0 \hbox{ citations} \qquad
2010:\ 10.2 \hbox{ citations} \qquad
2015:\ 6.3 \hbox{ citations.}
\end{equation}
There are certainly many reasons why these subgroups are not fully representative of their original groups. For example, all publications appearing in proceedings or in books are not linked to a JIF. Thus, for the next graph only the items of the list \eqref{eq_step_4} have been considered, but elsewhere let us stress that the more general list \eqref{eq_step_3} was preferred.

\begin{figure}[h]
\centering
\includegraphics[scale = 0.45]{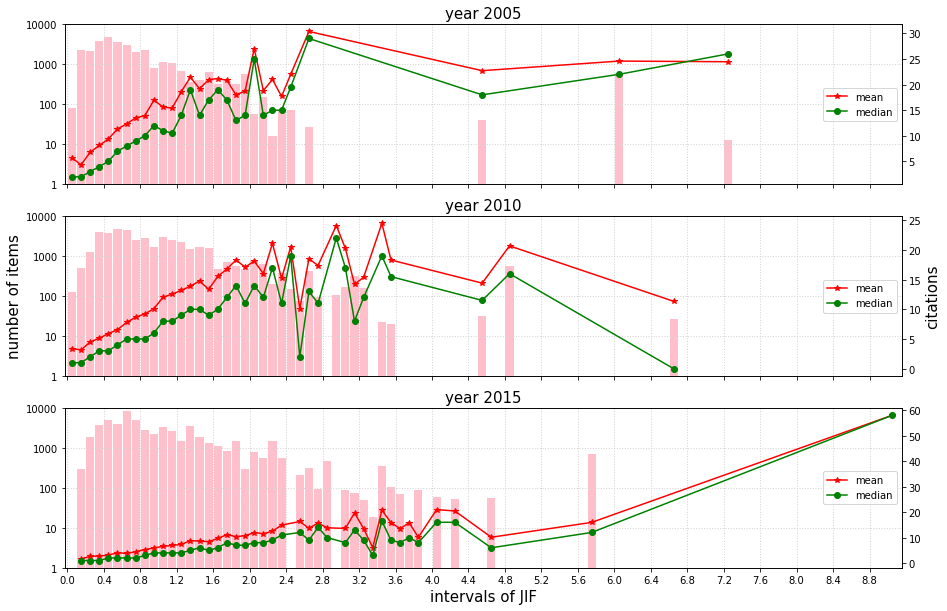}
\caption{Citations depending on the Journal Impact Factor}
\label{fig_JIF}
\end{figure}

In Figure \ref{fig_JIF} we report the citation means and the citation medians
as a function of the JIF. For the graphs, we have divided the possible values
of the JIF into subintervals of length $0.1$, collected all papers linked to a JIF in each subinterval and computed their mean and median. The information about the number of items corresponding to each subinterval is also mentioned. For a JIF below $2$, a nearly linear relation between JIF and citation mean or median is quite visible. On the other hand, for values of the JIF above $2$, the relation is less clear. But this part of the statistics is performed on much fewer items, and makes it less reliable.

\subsection{Research areas}

To each item, WoS associates one or several research area(s).
As mentioned in the previous section, our selection was based on
\emph{research area = mathematics}, which means that all data
have at least \emph{mathematics} in the list of their research areas.
However, most of them possess more than just one research area. For the data
from 2005, it turns out that only 12 additional research areas were contained
in at least $1\%$ of the considered items. The list of these research areas is:
\emph{Mathematics, Computer Science, Engineering,
Physics, Mechanics, Operations Research \& Management Science,
Mathematical \& Computational Biology, Mathematical Methods in Social Sciences,
Business \& Economics, Biochemistry \& Molecular Biology, Automation \& Control Systems, Science \& Technology, Biotechnology \& Applied Microbiology}.

Based on this list, we have computed the fraction, the mean and the median of all the items introduced in \eqref{eq_step_3} which possess one of these entries as a research area. The result is provided in Table \ref{t_ra}.

\begin{table}[h]
\begin{tabular}{|l|c|c|c|c|c|c|c|c|c|c|c|c|}
\hline
 & \multicolumn{4}{c|}{2005} & \multicolumn{4}{c|}{2010} & \multicolumn{4}{c|}{2015} \\ \hline
 & {\tiny \%} & {\tiny mean} & {\tiny rel. c.} & {\tiny med.} & {\tiny \%} & {\tiny mean} & {\tiny rel. c.} & {\tiny med. } & {\tiny \%} & {\tiny mean} & {\tiny rel. c.} & {\tiny med.} \\ \hline
{\tiny Mathematics} & 100 & 10.0 & 1.0 & 5 & 100 & 8.5 & 1.0 & 4 & 100 & 5.3 & 1.0 & 2 \\ \hline
{\tiny Comp. Science} & 15.2 & 9.9 & 1.0 & 4 & 15.0 & 8.4 & 1.0 & 3 & 9.9 & 6.7 & 1.3 & 3 \\ \hline
{\tiny Engineering} & 8.7 & 7.1 & 0.7 & 1 & 11.3 & 6.1 & 0.7 & 1 & 8.7 & 6.8 & 1.3 & 3 \\ \hline
{\tiny Physics} & 8.7 & 11.3 & 1.1 & 6 & 7.9 & 7.7 & 0.9 & 3 & 7.3 & 5.1 & 1.0 & 2 \\ \hline
{\tiny Mechanics} & 4.5 & 9.4 & 0.9 & 4 & 5.3 & 10.6 & 1.2 & 5 & 3.7 & 10.9 & 2.1 & 7 \\ \hline
{\tiny Op. Research} & 4.3 & 6.7 & 0.7 & 2 & 4.4 & 6.0 & 0.7 & 2 & 2.7 & 5.6 & 1.1 & 3 \\ \hline
{\tiny Math. \& C. Bio.} & 3.2 & 19.9 & 2.0 & 15 & 3.5 & 13.1 & 1.5 & 8 & 2.4 & 10.4 & 2.0 & 6 \\ \hline
{\tiny Math. M. Soc. S.} & 2.5 & 14.0 & 1.4 & 9 & 2.8 & 11.3 & 1.3 & 6 & 2.5 & 8.4 & 1.6 & 5 \\ \hline
{\tiny Bus. \& Eco.} & 2.8 & 8.7 & 0.9 & 1 & 2.3 & 9.1 & 1.1 & 4 & 2.2 & 6.4 & 1.2 & 3 \\ \hline
{\tiny Bio. \& M. Bio.} & 1.8 & 23.1 & 2.3 & 19 & 1.3 & 16.9 & 2.0 & 12 & 1.3 & 13.4 & 2.5 & 9 \\ \hline
{\tiny Auto. \& C. Syst.} & 1.7 & 11.5 & 1.1 & 6 & 1.8 & 11.9 & 1.4 & 6 & 1.8 & 12.0 & 2.3 & 8 \\ \hline
{\tiny Science \& Tech.} & 1.7 & 9.8 & 1.0 & 6 & 1.9 & 6.5 & 0.8 & 3 & 2.2 & 4.1 & 0.8 & 2 \\ \hline
{\tiny Bio. \& A. Mic.} & 1.6 & 23.9 & 2.4 & 20 & 1.1 & 17.4 & 2.1 & 13 & 1.0 & 14.8 & 2.8 & 11 \\ \hline
\end{tabular}
\caption{Citations for the main research areas}
\label{t_ra}
\end{table}

In this table, we have also reported the \emph{relative citation} which corresponds to the citation mean of a particular research area divided by the citation mean of the same year for all the data. It turns out that these computations give some interesting results: the range of this relative citation is between $0.7$ and $2.8$, with the lowest value shared in 2005 and 2010 by \emph{Engineering} and by \emph{Operations Research \& Management Science}. On the other hand, most of the highest values are reached by research areas related to biology or to biotechnology.
Note that the median follows a similar pattern, but since the computation of a relative median does not look so natural, we have refrained from providing such a relative information.
Thus, this table confirm that the citation mean or median really depend on the research areas, and that some striking differences exist. This fact is well documented, and has led to the developments of several relative indices, see for example \cite{AB, deB}.

\subsection{Categories}

Categories correspond to another indexation of the items chosen by WoS.
They correspond to rather broad research fields, but some of them coincide
also with research areas. The main list of categories (different from any research area) appearing in our items are the following:
\emph{Applied Mathematics,
Mathematics,
Statistics \& Probability,
Mathematics (Interdisciplinary Applications),
Computer Science (Interdisciplinary Applications),
Mathematical Physics,
Computer Science (Theory \& Methods),
Engineering (Multidisciplinary),
Mechanics}.

For each of these categories, the proportion, the citation mean, relative citation, and the median have been reported in Table \ref{t_cat}. The variations already observed in Table \ref{t_ra} are also visible here. Note that even in the three main categories, namely Mathematics, Applied mathematics, and Statistics and Probability, the means and the medians are clearly not equal, even if the variations are less pronounced than with categories further apart.

\begin{table}[h]
\begin{tabular}{|l|c|c|c|c|c|c|c|c|c|c|c|c|}
\hline
 & \multicolumn{4}{c|}{2005} & \multicolumn{4}{c|}{2010} & \multicolumn{4}{c|}{2015} \\ \hline
 & {\tiny \%} & {\tiny mean} & {\tiny rel. c.} & {\tiny med.} & {\tiny \%} & {\tiny mean} & {\tiny rel. c.} & {\tiny med. } & {\tiny \%} & {\tiny mean} & {\tiny rel. c.} & {\tiny med.} \\ \hline
{\tiny Math. app.} & 45.1 & 9.5 & 0.9 & 4 & 50.5 & 8.2 & 1.0 & 4 & 41.8 & 5.6 & 1.1 & 3 \\ \hline
{\tiny Mathematics} & 41.5 & 8.6 & 0.9 & 5 & 40.7 & 7.2 & 0.8 & 4 & 46.3 & 3.9 & 0.7 & 2 \\ \hline
{\tiny Sta. \& Prob.} & 16.2 & 12.8 & 1.3 & 7 & 15.4 & 10.5 & 1.2 & 6 & 15.8 & 6.5 & 1.2 & 3 \\ \hline
{\tiny Math., Int. App.} & 14.9 & 11.8 & 1.2 & 6 & 12.3 & 11.5 & 1.4 & 6 & 16.4 & 7.2 & 1.4 & 3 \\ \hline
{\tiny C. S., Int. App.} & 7.1 & 10.6 & 1.1 & 4 & 6.9 & 8.4 & 1.0 & 3 & 3.7 & 7.9 & 1.5 & 4 \\ \hline
{\tiny Math. Phys.} & 6.3 & 10.9 & 1.1 & 5 & 5.2 & 8.5 & 1.0 & 3 & 3.7 & 7.7 & 1.4 & 4 \\ \hline
{\tiny C. S., T. \& M.} & 5.0 & 9.6 & 1.0 & 4 & 5.0 & 6.8 & 0.8 & 3 & 3.8 & 4.7 & 0.9 & 2 \\ \hline
{\tiny Eng., Mult.} & 5.1 & 9.2 & 0.9 & 3 & 6.3 & 7.6 & 0.9 & 2 & 6.8 & 7.0 & 1.3 & 3 \\ \hline
{\tiny Mechanics} & 4.5 & 9.4 & 0.9 & 4 & 5.3 & 10.6 & 1.2 & 5 & 3.7 & 10.9 & 2.1 & 7 \\ \hline
\end{tabular}
\caption{Citations for the main categories}
\label{t_cat}
\end{table}

\section{Investigations with decision trees}\label{invest}

In the previous section, we studied the relationships between the number of citations and some individual predictors extracted from the WoS database. In several figures and tables, it clearly appears that the citations are related to these predictors. The problems left are how much information on the citations can be explained by the predictors and what is the relative importance among the predictors. To answer these questions, we use decision trees, as thoroughly introduced in \cite{BFOS}.

\subsection{Tree classifier}

\begin{figure}[h]
\centering
\includegraphics[scale = 0.36]{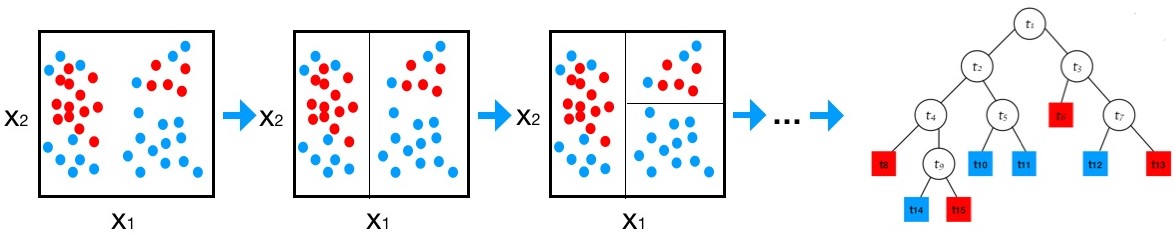}
\caption{Tree-structured classifier}
\label{fig:tr}
\end{figure}

A tree classifier is a procedure that divides a data set into two or more subsets based on some predetermined criteria. Let us explain this on a simple example. Suppose that the response contains two classes: Yes and No, which are represented by red and blue dots in Figure \ref{fig:tr}. Suppose also that there are two predictors $X_1$ and $X_2$ attached to each item. The tree classifier can identify the best split value of one of the predictors such that the purity in each subset is enhanced. After the splitting into two subsets, further splits will be carried out independently for each subset. The classifier goes through all the possible values of the predictors at each split. It stops splitting when a stopping rule is met. During the process, each subset is called a node, the nodes without further subsets are called leaves. When a node contains more than one class (which means that the node is not pure), the class with the majority of items is selected as the label class of the node.
The \emph{misclassification rate} of the node corresponds to the ratio of items in the node belonging to a class which is not the label class.

There exist several criteria for the choice of the split value,
but all of them are based on an \emph{impurity function} which has to be minimized.
More precisely, let $H$ denote such an impurity function, and let us consider a split of the content of one node $t$ into two subsets, called $t_{\emph{left}}$ and $t_{\emph{right}}$. Then one sets
\begin{equation}\label{eq_Q}
Q_t=\frac{N_{t_{\emph{left}}}}{N_{t}}H({t_{\emph{left}}}) + \frac{N_{t_{\emph{right}}}}{N_{t}}H(t_{\emph{right}}),
\end{equation}
where $N_{t}$, $N_{t_{\emph{left}}}$, and $N_{t_{\emph{right}}}$
denote the number of items in the note $t$ and in the two subsets.
The function $H$ is also evaluated on the items of $t_{\emph{left}}$ and $t_{\emph{right}}$.
By considering all possible splits, we choose the one which generates the minimal value for $Q_t$.

For the impurity function, a few canonical choices are possible. In order to define them, consider now that the
response contains $J$ classes, and assume that in a node $t$ the items are distributed following a distribution $\{p_j\}_{j=1}^J$, with $p_j$ the proportion of items in the class $j$. Then, some canonical impurity functions are
\begin{align*}
\mathrm{Gini\ index:} &\ \ H(t)=\sum_j p_j(1-p_j) \\
\mathrm{Entropy:} &\ \ H(t)=-\sum_j p_j \ln(p_j)\\
\mathrm{Misclassification:} &\ \ H(t)=1-\max_j(p_j)
\end{align*}
In the subsequent investigations, we shall use the impurity function provided
by the Gini index only.

By building a tree with the above process, we often end up with a very big tree: many leaves and a big height, which corresponds to the maximal distance between the first node (root) and the farthest leaf. Such a big tree leads often to an overfitting phenomenon. Thus, a pruning procedure need to be done to reserve the effective tree structure and to remove the risk of overfitting. One can perform the pruning procedure by removing successively the leaves with the least contribution of decreasing the impurity, namely the weakest leaves.
There exists several ways for implementing such a process, let us therefore only sketch the main ideas of the cost complexity pruning.
To each node $t$, one associates a real coefficient $\alpha_{eff}(t)$ which takes
into account the misclassification of the node, the misclassification of the subtree having $t$ as a root, and the number of leaves of the subtree, see for example \cite{Gini}.
Then, starting from $0$ and by slowly increasing a parameter $\alpha$, one prunes successively the nodes with $\alpha_{eff}$ smaller than $\alpha$.
Obviously, an additional stopping rule has to be fixed, otherwise the process would end up in keeping only the root, which means the original set of items without any subdivision. Again, several options exist.
Before presenting one of these stopping rules, let us mention some outcomes of the pruning operation.

Consider that $\alpha$ is slowly increasing from $0$, and that the pruning is taking place.
It is clear that the number of leaves and the height of the successive trees are decreasing. On the other hand, the total misclassification of the tree tends to increase.
Equivalently, the \emph{train accuracy} (ratio of correctly classified items in the leaves) tends to decrease as $\alpha$ increases.
It is thus natural to look for a suitable value of $\alpha$, leading neither to a too small tree which is not able to do effective classification, nor to a too large tree with high risk of overfitting.

\begin{figure}[h]
\centering
\includegraphics[scale = 0.25]{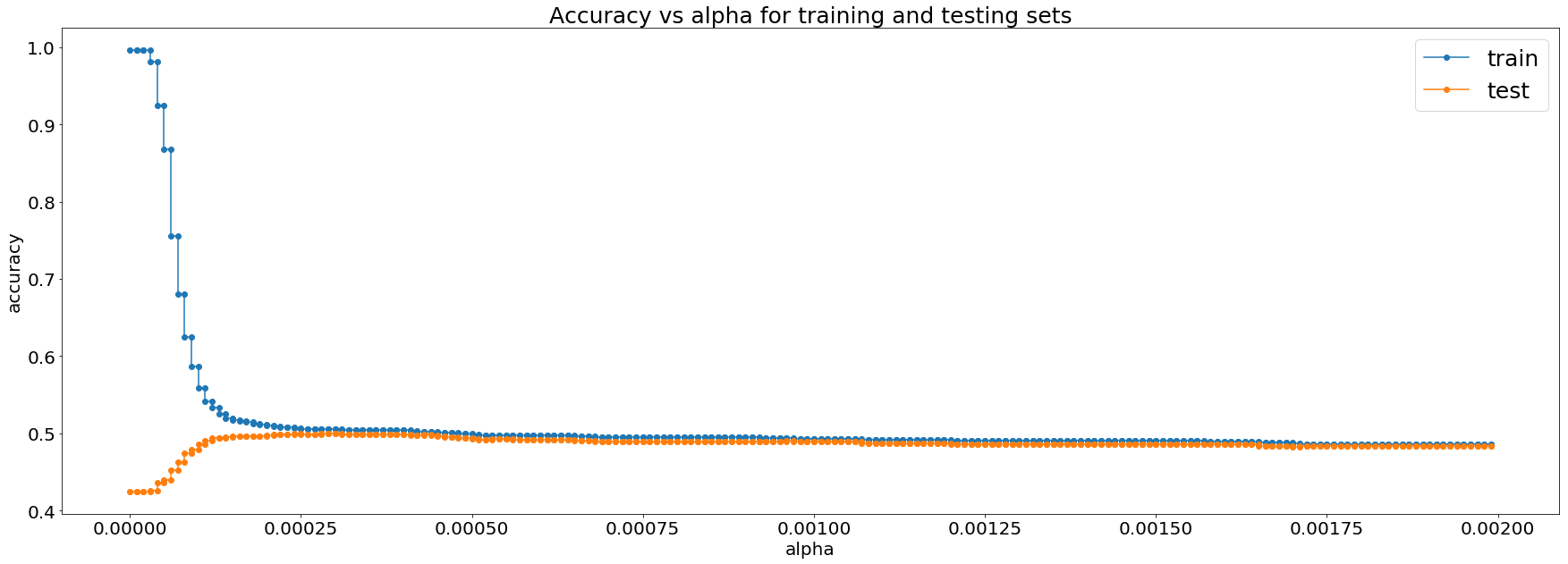}
\caption{Train accuracy and test accuracy, as a function of $\alpha$}
\label{alphas}
\end{figure}

Such a suitable $\alpha$ can be obtained by testing the tree on new items. Indeed,
consider a new item having all necessary predictors and labeled by one class.
According to the value of its predictors, the item can be placed in a unique leaf with a labeled class. It may be correctly or incorrectly classified. By repeating this operation on several new items, one get a \emph{test accuracy}, the ratio of correctly classified new items.
Again, if one considers the family of trees obtained by pruning according to the parameter $\alpha$, one observes that the test accuracy starts by increasing with $\alpha$, before decreasing again. Since we are interested in the highest test accuracy, we then select the optimal value $\alpha_{opt}$ of the parameter $\alpha$ corresponding to the maximum test accuracy. A typical example of the train accuracy and the test accuracy as a function of $\alpha$ is provided in Figure~\ref{alphas}.
It is then this $\alpha_{opt}$ which is used for stopping the pruning process.

\subsection{The experiment}\label{c_t}

In this section, we implement these processes and describe the precise experiment we have performed.
The analyze tool package is provided by Scikit-learn (Supervised Learning / Decision Trees), see \cite{PVG}.
The experiments will be done independently on the three datasets of 2005, 2010, and 2015.
For the predictors, we shall use those introduced and discussed in Section \ref{sec_pred}.
However, for the predictor $jif$, some items miss this information because they do not have an associated JIF. We have then decided to perform the experiment independently on two lists of items. The first list contains all items, as shown in \eqref{eq_step_3}, and the predictor $jif$ will not be used. The second list contains items with a JIF, as shown in \eqref{eq_step_4}, and the predictor $jif$ is included in the list of all possible predictors.

For the six lists of items mentioned above, we define a family of classes, namely a partition of the set $\{0,1,2,\ldots,63\}$, corresponding to the $J$ classes mentioned in the previous section. The partitions should be relevant and understandable and they should also be chosen accordingly to the specificity of the different lists of items.
For example, the following partition will be used for the first list of items of 2005:
\begin{equation}\label{eq_classes_2005}
\hbox{weak:}\ [0,5],\quad
\hbox{normal:}\ [6,12], \quad
\hbox{good:}\ [13,20], \quad
\hbox{very good:}\ [21,40], \quad
\hbox{excellent:}\ [41, 63].
\end{equation}

Let us now describe the precise construction of the tree classifier, and the pruning process.

\begin{enumerate}
\item[(i)] Fix one dataset among the six presented in \eqref{eq_step_3} and in \eqref{eq_step_4},
\item[(ii)] Fix one relevant partition for the dataset, as for example the one presented in \eqref{eq_classes_2005}. The number of classes defined by this partition is denoted by $J$ and corresponds to the number of intervals,
\item[(iii)] Label the items in the dataset with one of the $J$ classes, according to their citations, and select randomly $X$ items for each class. For our experiment we have
chosen $X$ equal to $80\%$ of the items of the smallest class (which has always been the one corresponding to the highest citations class). One ends up with $JX$ items equally distributed among the $J$ classes,
\item[(iv)] Divide randomly these $JX$ items into $K$ folds $\{\Lambda_k\}_{k=1}^K$ of equal size, for $K\in \N$, and fix $k=1$,
\item[(v)] The fold $\Lambda_k$ is called the \emph{test set}, and the others $K-1$ folds are combined in a \emph{training set}. A classification tree is build with the training set,
\item[(vi)] On the classification tree, the pruning process introduced in the previous section is performed: the computation of $\alpha_{eff}(t)$ for each node $t$, the increase of $\alpha$, the pruning of the weakest node (\emph{i.e.}~the ones with the smallest $\alpha_{eff}$), the computation of the train accuracy and of the test accuracy. These accuracies, denoted respectively by $a_k(\alpha)$ and $b_k(\alpha)$, are computed for each $\alpha$ (but are piecewise constant) and are stored. The value $k$ is updated by setting $k:=k+1$,
\item[(vii)] The steps (v) and (vi) are repeated as long as $k\leq K$,
\item[(viii)] For each $\alpha$, the average $a(\alpha)$ and average $b(\alpha)$ are computed by averaging the $K$ values and $\alpha_{opt}$ is deduced by
generating a graph like Figure \ref{alphas} with the average values ($K$-fold cross-validation),
\item[(ix)] The optimal tree is built based on the $JX$ items and pruned with the $\alpha_{opt}$ found at the previous step.
\end{enumerate}

Note that in our experiments we have used the parameter $K=5$.
Once this optimal tree is built, the relative importance of the predictors can be
obtained. Indeed, the relative importance of the predictors can be computed by looking at the \emph{weighted impurity decrease} at each node. By using the notation already introduced in \eqref{eq_Q}, this quantity is provided for each node $t$ by the formula
$$
\frac{N_{t}}{N} \Big(H(t)-\frac{N_{t_{\emph{left}}}}{N_{t}}H({t_{\emph{left}}}) - \frac{N_{t_{\emph{right}}}}{N_{t}}H(t_{\emph{right}})\Big),
$$
where $N$ is the total number of items in the tree.
Clearly, if a node is a leaf, there is no contribution to be subtracted.
The above quantity provides an information about the decay in the impurity provided by a subdivision. Then, since each subdivision is associated with a single predictor, this decay in the impurity is gained by the corresponding predictor.
By summing up the decay in impurity due to all subdivisions
associated with one predictor, one obtains the total decay in impurity due to this predictor. The predictors are finally ordered by the decreasing order of their
total decay in impurity.

In Table \ref{t_tree}, we provide various information obtained with the tree classifier, namely the relative importance of the predictors, as mentioned above, but also
the number of leaves and the height of each tree. The average train accuracy and the test accuracy are also reported. These two values correspond to the two accuracies obtained at $\alpha_{opt}$ after cross-validation.

\begin{table}[h]
\centering
\begin{tabular}{|l|c|c|c|c|c|c|}
\hline
 & \multicolumn{2}{c|}{2005} & \multicolumn{2}{c|}{2010} & \multicolumn{2}{c|}{2015} \\ \hline
JIF & No & Yes & No & Yes & No & Yes \\ \hline
classes & \begin{tabular}[c]{@{}c@{}}{[}0,5{]}\\ {[}6,12{]}\\ {[}13,20{]}\\ {[}21,40{]}\\ {[}41,63{]}\end{tabular} & \begin{tabular}[c]{@{}c@{}}{[}0,6{]}\\ {[}7,14{]}\\ {[}15,23{]}\\ {[}24,45{]}\\ {[}46,63{]}\end{tabular} & \begin{tabular}[c]{@{}c@{}}{[}0,4{]}\\ {[}5,10{]}\\ {[}11,20{]}\\ {[}21,63{]}\end{tabular} & \begin{tabular}[c]{@{}c@{}}{[}0,5{]}\\ {[}6,12{]}\\ {[}13,24{]}\\ {[}25,63{]}\end{tabular} & \begin{tabular}[c]{@{}c@{}}{[}0,3{]}\\ {[}4,9{]}\\ {[}10,63{]}\end{tabular} & \begin{tabular}[c]{@{}c@{}}{[}0,4{]}\\ {[}5,11{]}\\ {[}12,63{]}\end{tabular} \\ \hline
\# leaves & 24 & 15 & 27 & 29 & 49 & 25 \\ \hline
height & 7 & 6 & 8 & 6 & 10 & 6 \\ \hline
av. train acc. (\%) & 32.9 & 34.5 & 39.3 & 40 & 51.3 & 51.5 \\ \hline
av. test acc. (\%) & 31.3 & 30 & 38.4 & 37.8 & 50.4 & 50.7 \\ \hline
\emph{references} & 1 & 2 & 1 & 2 & 1 & 2 \\ \hline
\emph{authors} & & 4 & 3 & 3 & 3 & 4 \\ \hline
\emph{countries} & & & & & & 3 \\ \hline
\emph{institutes} & 2 & & & & & \\ \hline
\emph{pages} & 4 & 3 & 4 & 4 & 2 & \\ \hline
\emph{jif} & \cellcolor[HTML]{333333}{\color[HTML]{333333} } & 1 & \cellcolor[HTML]{333333} & 1 & \cellcolor[HTML]{333333} & 1 \\ \hline
\emph{research area} & 3 {\tiny (ENG.)} & & 2 {\tiny (ENG.)} & & 4 {\tiny (Bio. \& A. M.)} & \\ \hline
Experiment (\%) & 39 & 40 & 44 & 43 & 53 & 53 \\ \hline
\end{tabular}
\caption{Experiments with tree classifier}
\label{t_tree}
\end{table}

Let us now make several comments about Table \ref{t_tree}.

\noindent
(i) First of all, the number of classes and their precise values is partially arbitrary, and was determined after several preliminary tests. Note that we considered less classes for the recent data since the citations accumulate with time.

\noindent
(ii) The size of the resulting tree is determined by the computation of $\alpha_{opt}$
as explained above. We were surprised that these trees are relatively small, with a number of leaves between 15 and 49, and a height of maximum 10. On the other hand, the 6 trees contained between 4'800 and 20'000 items.

\noindent
(iii) About the accuracies: since the $J$ classes are of equal size, a random guess for an item of the test set would give a correct prediction with a probability of $20\%$ for the data of 2005, $25\%$ for the data of 2010, and $33.3\%$ for the data of 2015. Thus, the difference between these values, and the test accuracy can be understood as the gain in prediction due to the tree. However, since the train set and the test set do not follow the initial distributions of items, a different computation has to be performed for an arbitrary set of items in the initial dataset (see below).

\noindent
(iv) The ranking of the predictors for each of the 6 trees is reported in the table, but only for the first 4. In turns out that only 7 predictors among 10 appear in the ranking. For publications which appear in a \emph{jif}ted journal, this predictor is always chosen first, and the number of references used by the authors is chosen as the second predictor. On the other hand, for the larger set of publications, with no use of the predictor \emph{jif}, it appears that the number of references is always chosen
as the first predictor, while the second one is different in the three experiments.
Note that the importance of the predictor \emph{references} was already
anticipated in Section \ref{sec_ind_pre} just by looking at the sharp contrasts
of Figure \ref{fig_references}.

\noindent
(v) An additional experiment has been performed with the items not used for the construction of the trees. Indeed, the $20\%$ of the smallest class was still available, and $20\%$ of the untouched items in the other classes could be selected randomly.
We apply the classification tree on the new test set which is made up of these items. The last row of Table \ref{t_tree} provides the
percentage of the correctly assigned classes. Clearly, these accuracies do not coincided
with the train accuracy or the test accuracy, since the corresponding items do not share the same distributions. The interpretation is the following: given an arbitrary item from one of the three years, and based on the constructed trees, our ability to predict correctly the citation class corresponds to the last row of the table. Not surprisingly, this accuracy increases as the number of classes decreases, but the knowledge of the JIF for the item does not improve our prediction.
Even if these numbers show the limitations of our approach, it also provides a heart-warming message: the content of a publication still matters for the citations, and any
bibliometric analysis won't be able to predict this.

\section{About countries}\label{sec_countries}

In this last section we provide some statistics related to countries. Indeed, by selecting the items according to the location of the corresponding research institutes, some additional information can be extracted.

In Table \ref{t_different_countries} we provide a comparison between the main 25 countries (according to their number of publications). These statistics can be thought as a kind of relative performance. More precisely, Table \ref{t_different_countries} contains the following entries:
\begin{enumerate}
\item[$\%:$] Percentage of the publications having at least one author from the given country,
\item[h.c.:] (highly cited) Percentage of publications having more than 63 citations with
 at least one author from the given country,
\item[mn:] (mean) Citation mean for the publications having less than 64 citations and at least one authors from the given country,
\item[md:] (median) Median for the publications having less than 64 citations and at least one authors from the given country,
\item[r.c.:] (relative citation) Ratio of mn by the citation mean over all publications with less than 64 citations.
\end{enumerate}
The following abbreviations for the countries is used:
US = USA, CN = People's R. China, FR = France,
DE = Germany, IT = Italy, UK = England, CA = Canada,
JP = Japan, ES = Spain, RU = Russia, AU = Australia,
KR	= South Korea, PL = Poland, IL = Israel, NL = Netherlands,
IN = India, BR = Brazil, TW = Taiwan, BE = Belgium,
CH	= Switzerland, SE = Sweden, GR = Greece, CZ = Czech Republic,
AT	 = Austria, TR = Turkey.

\begin{table}[h]
\centering
\scalebox{0.8}{
\begin{tabular}{|l|c|c|c|c|c|c|c|c|c|c|c|c|c|c|c|}
\hline
 & \multicolumn{5}{c|}{2005} & \multicolumn{5}{c|}{2010} & \multicolumn{5}{c|}{2015} \\ \hline
 & \% & h.c. & mn & md & r.c. & \% & h.c. & mn & md & r.c. & \% & h.c. & mn & md & r.c. \\ \hline
US & 27.7 & 40.6 & 12.0 & 7 & 1.2 & 24.0 & 37.3 & 10.1 & 5 & 1.2 & 20.7 & 32.9 & 6.5 & 3 & 1.2 \\ \hline
CN & 10.8 & 12.8 & 11.0 & 6 & 1.1 & 17.1 & 17.2 & 8.6 & 4 & 1.0 & 19.1 & 29.5 & 6.5 & 3 & 1.2 \\ \hline
FR & 8.0 & 9.3 & 12.0 & 7 & 1.2 & 7.7 & 7.1 & 10.2 & 6 & 1.2 & 7.0 & 7.4 & 6.0 & 3 & 1.1 \\ \hline
DE & 7.3 & 8.7 & 11.2 & 6 & 1.1 & 7.3 & 7.7 & 8.9 & 5 & 1.0 & 6.9 & 10.6 & 6.1 & 3 & 1.2 \\ \hline
IT & 5.5 & 4.8 & 9.9 & 5 & 1.0 & 5.1 & 4.2 & 9.4 & 5 & 1.1 & 5.1 & 7.4 & 6.6 & 4 & 1.3 \\ \hline
UK & 5.3 & 7.8 & 12.2 & 7 & 1.2 & 5.0 & 8.1 & 10.7 & 6 & 1.3 & 4.7 & 11.1 & 6.6 & 4 & 1.2 \\ \hline
CA & 4.6 & 5.0 & 11.2 & 6 & 1.1 & 4.2 & 5.2 & 9.8 & 5 & 1.2 & 3.5 & 3.7 & 5.7 & 3 & 1.1 \\ \hline
JP & 4.4 & 2.5 & 8.5 & 4 & 0.8 & 4.3 & 2.1 & 6.4 & 3 & 0.8 & 3.7 & 2.7 & 4.2 & 2 & 0.8 \\ \hline
ES & 4.1 & 3.4 & 11.0 & 6 & 1.1 & 4.3 & 4.2 & 8.9 & 5 & 1.1 & 3.7 & 3.2 & 5.8 & 3 & 1.1 \\ \hline
RU & 4.1 & 1.1 & 6.2 & 2 & 0.6 & 3.8 & 0.9 & 5.6 & 2 & 0.7 & 4.8 & 1.4 & 3.8 & 2 & 0.7 \\ \hline
AU & 2.9 & 3.3 & 9.7 & 4 & 1.0 & 2.1 & 3.1 & 9.7 & 5 & 1.1 & 2.5 & 4.0 & 6.1 & 3 & 1.2 \\ \hline
KR & 2.2 & 1.5 & 9.1 & 4 & 0.9 & 2.0 & 1.9 & 8.3 & 4 & 1.0 & 2.5 & 2.6 & 4.9 & 2 & 0.9 \\ \hline
PL & 2.1 & 0.7 & 8.7 & 4 & 0.9 & 2.2 & 0.8 & 6.9 & 3 & 0.8 & 2.6 & 1.8 & 4.7 & 2 & 0.9 \\ \hline
IL & 1.9 & 1.9 & 11.2 & 6 & 1.1 & 1.7 & 1.6 & 8.7 & 5 & 1.0 & 1.4 & 1.4 & 5.5 & 3 & 1.0 \\ \hline
NL & 1.8 & 2.1 & 11.6 & 6 & 1.2 & 1.7 & 2.3 & 9.8 & 6 & 1.2 & 1.3 & 1.6 & 6.6 & 3 & 1.3 \\ \hline
IN & 1.7 & 1.3 & 10.4 & 6 & 1.0 & 2.3 & 2.3 & 9.1 & 4 & 1.1 & 4.1 & 2.7 & 4.5 & 2 & 0.8 \\ \hline
BR & 1.6 & 1.3 & 10.3 & 6 & 1.0 & 1.8 & 1.6 & 8.7 & 5 & 1.0 & 2.1 & 0.8 & 5.0 & 3 & 1.0 \\ \hline
TW & 1.6 & 1.6 & 12.4 & 7 & 1.2 & 1.8 & 1.5 & 9.3 & 5 & 1.1 & 1.4 & 0.6 & 4.8 & 2 & 0.9 \\ \hline
BE & 1.4 & 1.9 & 11.5 & 6 & 1.1 & 1.3 & 1.3 & 9.8 & 5 & 1.2 & 1.1 & 1.9 & 6.9 & 4 & 1.3 \\ \hline
CH & 1.3 & 2.1 & 12.9 & 7 & 1.3 & 1.2 & 2.4 & 10.8 & 6 & 1.3 & 1.4 & 2.7 & 7.9 & 5 & 1.5 \\ \hline
SE & 1.2 & 1.2 & 10.8 & 6 & 1.1 & 1.1 & 0.8 & 7.9 & 4 & 0.9 & 1.1 & 1.6 & 5.4 & 2 & 1.0 \\ \hline
GR & 1.2 & 1.0 & 9.1 & 4 & 0.9 & 0.9 & 0.6 & 8.5 & 5 & 1.0 & 0.7 & 1.3 & 5.6 & 2 & 1.1 \\ \hline
CZ & 1.2 & 0.6 & 7.5 & 2 & 0.7 & 1.3 & 0.6 & 6.9 & 3 & 0.8 & 1.5 & 0.0 & 3.8 & 1 & 0.7 \\ \hline
AT & 1.1 & 1.5 & 10.4 & 5 & 1.0 & 1.2 & 1.8 & 9.7 & 5 & 1.1 & 1.2 & 2.3 & 6.1 & 3 & 1.1 \\ \hline
TR & 1.1 & 0.9 & 10.5 & 6 & 1.0 & 1.7 & 2.0 & 9.9 & 5 & 1.2 & 2.0 & 1.4 & 5.0 & 2 & 0.9 \\ \hline
\end{tabular}}
\caption{Statistics for the main countries}
\label{t_different_countries}
\end{table}

Let us make some comments about this table. First of all, columns $\%$ and h.c. sum up to more than 100 (when all countries are considered) because publications involving authors from different countries are counted more than once. Note also that it is the first (and last) time that outliers (namely publications with more than 63 citations) are used:
Columns h.c. provides the information on how much the countries are involved in the highly cited publications.
The columns with r.c. can be thought as a comparison between the performance of these countries: it is rather striking that the relative citations take values between 0.6 and 1.5. Some countries are clearly performing better than others. On the other hand, these data do not present a clear pattern over the 3 years considered, and the relative citations are quite stable. Only a few countries have a small variation of their relative citations over the 3 years, but the evolution is not so noticeable. Most probably, a period of 10 years is not long enough to really assert a real change of performance for a given country.

The next table, Table \ref{t_collaboration_countries}, is more related to the importance of international collaborations for each country. By international collaboration we mean a publication with one author from the given country, and at least one author from another country. More precisely, Table \ref{t_collaboration_countries} contains the following entries, again for the main 25 countries:
\begin{enumerate}
\item[$\%:$] Percentage of publications of a given country which are international collaborations,
\item[mn:] (mean) Citation mean for the publications of a given country which are international collaborations and which have less than 64 citations,
\item[md:] (median) Median for the publications of a given country which are international collaborations and which have less than 64 citations
\item[rcc:] (relative citation for international collaborations) Ratio of mn by the citation mean over all publications of this country with less than 64 citations,
\item[rc2:] (relative citations for international collaborations / 2 authors) Ratio of mn by the citation mean over all publications of this country with less than 64 citations and at least 2 authors.
\end{enumerate}

Note that we have added the column rc2 in order to eliminate a bias. Indeed, an international collaboration involves at least 2 authors (with a few exceptions already mentioned in Section \ref{subsection_nb_countries}) while arbitrary publications from any country also include many single author productions. Since these publications are usually less cited (see Table \ref{t_authors_2}), we eliminate this bias by considering only publications with at least 2 authors.

\begin{table}[h]
\centering
\scalebox{0.8}{
\begin{tabular}{|l|c|c|c|c|c|c|c|c|c|c|c|c|c|c|c|}
\hline
 & \multicolumn{5}{c|}{2005} & \multicolumn{5}{c|}{2010} & \multicolumn{5}{c|}{2015} \\ \hline
 & \% & mn & md & rcc & rc2 & \% & mn & md & rcc & rc2 & \% & mn & md & rcc & rc2 \\ \hline
US & 30.1 & 14.4 & 9 & 1.2 & 1.0 & 34.9 & 12.3 & 7 & 1.2 & 1.0 & 42.4 & 7.5 & 4 & 1.2 & 1.0 \\ \hline
CN & 23.7 & 14.1 & 9 & 1.3 & 1.2 & 21.2 & 12.4 & 7 & 1.4 & 1.4 & 25.5 & 8.6 & 5 & 1.3 & 1.3 \\ \hline
FR & 38.7 & 14.4 & 9 & 1.2 & 1.1 & 48.1 & 11.5 & 7 & 1.1 & 1.0 & 54.7 & 6.6 & 4 & 1.1 & 1.0 \\ \hline
DE & 40.1 & 13.6 & 9 & 1.2 & 1.1 & 47.6 & 10.8 & 7 & 1.2 & 1.1 & 51.8 & 7.1 & 4 & 1.2 & 1.0 \\ \hline
IT & 33.7 & 13.6 & 8 & 1.4 & 1.2 & 42.4 & 11.5 & 7 & 1.2 & 1.1 & 49.5 & 7.8 & 5 & 1.2 & 1.1 \\ \hline
UK & 46.8 & 14.3 & 9 & 1.2 & 1.0 & 54.0 & 12.3 & 7 & 1.1 & 1.0 & 61.5 & 7.3 & 4 & 1.1 & 1.0 \\ \hline
CA & 51.2 & 13.6 & 8 & 1.2 & 1.1 & 53.7 & 11.4 & 7 & 1.2 & 1.1 & 60.8 & 6.4 & 3 & 1.1 & 1.0 \\ \hline
JP & 23.3 & 12.3 & 8 & 1.4 & 1.2 & 30.3 & 8.7 & 5 & 1.4 & 1.2 & 33.3 & 6.3 & 3 & 1.5 & 1.3 \\ \hline
ES & 36.0 & 14.2 & 10 & 1.3 & 1.2 & 46.2 & 10.5 & 6 & 1.2 & 1.1 & 54.4 & 6.4 & 4 & 1.1 & 1.1 \\ \hline
RU & 29.3 & 11.5 & 7 & 1.8 & 1.5 & 30.3 & 9.0 & 5 & 1.6 & 1.2 & 25.0 & 5.9 & 3 & 1.5 & 1.3 \\ \hline
AU & 40.9 & 14.2 & 9 & 1.5 & 1.4 & 56.8 & 11.2 & 6 & 1.2 & 1.1 & 57.7 & 8.2 & 4 & 1.3 & 1.2 \\ \hline
KR & 31.5 & 13.0 & 7 & 1.4 & 1.3 & 41.9 & 10.7 & 6 & 1.3 & 1.2 & 40.7 & 6.8 & 3 & 1.4 & 1.2 \\ \hline
PL & 32.0 & 11.9 & 8 & 1.4 & 1.2 & 34.1 & 9.3 & 5 & 1.3 & 1.2 & 38.1 & 6.1 & 3 & 1.3 & 1.1 \\ \hline
IL & 46.9 & 13.2 & 8 & 1.2 & 1.0 & 54.0 & 10.4 & 6.5 & 1.2 & 1.1 & 58.1 & 6.7 & 4 & 1.2 & 1.1 \\ \hline
NL & 42.6 & 13.9 & 9 & 1.2 & 1.1 & 54.9 & 11.5 & 7 & 1.2 & 1.1 & 63.7 & 6.8 & 3 & 1.0 & 1.0 \\ \hline
IN & 31.0 & 12.6 & 7 & 1.2 & 1.2 & 31.3 & 10.7 & 6 & 1.2 & 1.1 & 26.4 & 6.3 & 3 & 1.4 & 1.3 \\ \hline
BR & 39.8 & 13.4 & 9 & 1.3 & 1.2 & 44.1 & 10.2 & 6 & 1.2 & 1.1 & 43.8 & 6.6 & 4 & 1.3 & 1.3 \\ \hline
TW & 29.5 & 13.6 & 9 & 1.1 & 1.1 & 35.4 & 10.8 & 7 & 1.2 & 1.1 & 40.8 & 6.4 & 4 & 1.3 & 1.3 \\ \hline
BE & 45.1 & 14.5 & 10 & 1.3 & 1.1 & 57.4 & 11.1 & 6 & 1.1 & 1.1 & 66.9 & 7.1 & 4 & 1.0 & 1.0 \\ \hline
CH & 49.2 & 17.4 & 13 & 1.3 & 1.2 & 61.7 & 11.8 & 8 & 1.1 & 1.0 & 68.6 & 8.6 & 5 & 1.1 & 1.0 \\ \hline
SE & 44.6 & 12.8 & 7 & 1.2 & 1.1 & 50.9 & 9.8 & 5 & 1.2 & 1.0 & 62.2 & 6.3 & 3 & 1.2 & 1.1 \\ \hline
GR & 27.5 & 14.7 & 10 & 1.6 & 1.5 & 45.3 & 9.9 & 5 & 1.2 & 1.1 & 54.9 & 7.4 & 3 & 1.3 & 1.2 \\ \hline
CZ & 37.1 & 12.5 & 7 & 1.7 & 1.4 & 44.8 & 9.8 & 6 & 1.4 & 1.2 & 39.4 & 6.2 & 3 & 1.6 & 1.5 \\ \hline
AT & 40.9 & 14.4 & 10 & 1.4 & 1.2 & 59.5 & 10.7 & 6 & 1.1 & 1.0 & 62.2 & 6.7 & 4 & 1.1 & 1.0 \\ \hline
TR & 21.9 & 11.8 & 8 & 1.1 & 1.1 & 25.6 & 11.7 & 6 & 1.2 & 1.1 & 33.5 & 6.9 & 4 & 1.4 & 1.3 \\ \hline
\end{tabular}}
\caption{Statistics about international collaborations for the main countries}
\label{t_collaboration_countries}
\end{table}

By looking at this table, the interest of international collaborations is quite clear. All the relative citations appearing in rcc or rc2 are bigger or equal to 1. In fact, some countries benefit a lot from international collaborations, having a factor rc2 taking a maximum value of 1.5. On the other hand, for some countries, publications which are international collaborations do not receive substantially more citations than publications involving only researchers in this country.
As a rule and not surprisingly, authors in a country with a low relative citation in Table \ref{t_different_countries} benefit more from international collaborations than authors from a country with a high relative citation. Fortunately, no researcher from any country is penalized by establishing international collaboration: such an unfortunate situation
would end up with a rcc or rc2 smaller than 1 in Table \ref{t_collaboration_countries}.

\begin{table}
\centering
\scalebox{0.75}{
\begin{tabular}{|l|c|c|c|c|c|c|c|c|c|c|c|c|c|c|c|c|}
\hline
 & US & CN & FR & DE & IT & UK & CA & JP & ES & RU & AU & KR & PL & IL & NL & IN \\ \hline
US & & \begin{tabular}[c]{@{}l@{}}36.7\\ 37.5\\ 40.1\end{tabular} & \begin{tabular}[c]{@{}l@{}}23.9\\ 21.5\\ 18.8\end{tabular} & \begin{tabular}[c]{@{}l@{}}24.2\\ 22.8\\ 22.7\end{tabular} & \begin{tabular}[c]{@{}l@{}}21.6\\ 22.2\\ 19.9\end{tabular} & \begin{tabular}[c]{@{}l@{}}27.9\\ 22.3\\ 25\end{tabular} & \begin{tabular}[c]{@{}l@{}}37\\ 34.6\\ 32.7\end{tabular} & \begin{tabular}[c]{@{}l@{}}27.2\\ 18.5\\ 22.3\end{tabular} & \begin{tabular}[c]{@{}l@{}}25.1\\ 16.6\\ 15.6\end{tabular} & \begin{tabular}[c]{@{}l@{}}20.7\\ 19.3\\ 18.8\end{tabular} & \begin{tabular}[c]{@{}l@{}}26.9\\ 19.7\\ 19.5\end{tabular} & \begin{tabular}[c]{@{}l@{}}46.9\\ 38.6\\ 33.6\end{tabular} & \begin{tabular}[c]{@{}l@{}}29\\ 20.5\\ 15\end{tabular} & \begin{tabular}[c]{@{}l@{}}54.5\\ 46.3\\ 49.2\end{tabular} & \begin{tabular}[c]{@{}l@{}}28.3\\ 23.6\\ 24.2\end{tabular} & \begin{tabular}[c]{@{}l@{}}36.1\\ 23\\ 22.2\end{tabular} \\ \hline
CN & \begin{tabular}[c]{@{}l@{}}11.3\\ 16.1\\ 22.2\end{tabular} & & \begin{tabular}[c]{@{}l@{}}3.7\\ 5.6\\ 6.7\end{tabular} & \begin{tabular}[c]{@{}l@{}}5.3\\ 5.3\\ 6.8\end{tabular} & \begin{tabular}[c]{@{}l@{}}3.3\\ 3.5\\ 3.1\end{tabular} & \begin{tabular}[c]{@{}l@{}}6.4\\ 7.7\\ 10.3\end{tabular} & \begin{tabular}[c]{@{}l@{}}13.5\\ 13\\ 18.7\end{tabular} & \begin{tabular}[c]{@{}l@{}}13.1\\ 18\\ 16.7\end{tabular} & \begin{tabular}[c]{@{}l@{}}2.2\\ 2.9\\ 4.2\end{tabular} & \begin{tabular}[c]{@{}l@{}}0.4\\ 3.1\\ 4.1\end{tabular} & \begin{tabular}[c]{@{}l@{}}15.2\\ 22\\ 26.8\end{tabular} & \begin{tabular}[c]{@{}l@{}}15.3\\ 21.6\\ 19.6\end{tabular} & \begin{tabular}[c]{@{}l@{}}2.3\\ 2.3\\ 5.4\end{tabular} & \begin{tabular}[c]{@{}l@{}}2.4\\ 2.1\\ 6\end{tabular} & \begin{tabular}[c]{@{}l@{}}1.7\\ 3.6\\ 5.6\end{tabular} & \begin{tabular}[c]{@{}l@{}}4.1\\ 6.5\\ 8.3\end{tabular} \\ \hline
FR & \begin{tabular}[c]{@{}l@{}}8.8\\ 9.5\\ 8.2\end{tabular} & \begin{tabular}[c]{@{}l@{}}4.5\\ 5.7\\ 5.3\end{tabular} & & \begin{tabular}[c]{@{}l@{}}8.2\\ 10\\ 9.9\end{tabular} & \begin{tabular}[c]{@{}l@{}}16.8\\ 17.9\\ 15.2\end{tabular} & \begin{tabular}[c]{@{}l@{}}7.2\\ 8.8\\ 9.6\end{tabular} & \begin{tabular}[c]{@{}l@{}}8\\ 9.3\\ 10\end{tabular} & \begin{tabular}[c]{@{}l@{}}6.1\\ 6.7\\ 9.5\end{tabular} & \begin{tabular}[c]{@{}l@{}}12.5\\ 12.4\\ 11.7\end{tabular} & \begin{tabular}[c]{@{}l@{}}13.8\\ 13.8\\ 11.4\end{tabular} & \begin{tabular}[c]{@{}l@{}}3.2\\ 7.7\\ 4.8\end{tabular} & \begin{tabular}[c]{@{}l@{}}3.6\\ 5.1\\ 4.2\end{tabular} & \begin{tabular}[c]{@{}l@{}}12.7\\ 10.2\\ 10.8\end{tabular} & \begin{tabular}[c]{@{}l@{}}8\\ 10.1\\ 8.7\end{tabular} & \begin{tabular}[c]{@{}l@{}}7.9\\ 7.1\\ 9.9\end{tabular} & \begin{tabular}[c]{@{}l@{}}7\\ 5.1\\ 4.9\end{tabular} \\ \hline
DE & \begin{tabular}[c]{@{}l@{}}8.4\\ 9.4\\ 9.3\end{tabular} & \begin{tabular}[c]{@{}l@{}}6.1\\ 5.1\\ 5\end{tabular} & \begin{tabular}[c]{@{}l@{}}7.8\\ 9.3\\ 9.2\end{tabular} & & \begin{tabular}[c]{@{}l@{}}11.5\\ 11.7\\ 12.1\end{tabular} & \begin{tabular}[c]{@{}l@{}}10.1\\ 11.7\\ 11.6\end{tabular} & \begin{tabular}[c]{@{}l@{}}5.4\\ 5.5\\ 6.2\end{tabular} & \begin{tabular}[c]{@{}l@{}}13.1\\ 8.5\\ 8.8\end{tabular} & \begin{tabular}[c]{@{}l@{}}5.8\\ 8.2\\ 7.5\end{tabular} & \begin{tabular}[c]{@{}l@{}}14.6\\ 12.6\\ 12.2\end{tabular} & \begin{tabular}[c]{@{}l@{}}6.1\\ 8.8\\ 6.1\end{tabular} & \begin{tabular}[c]{@{}l@{}}3.9\\ 4\\ 5.9\end{tabular} & \begin{tabular}[c]{@{}l@{}}13.7\\ 14.7\\ 12.9\end{tabular} & \begin{tabular}[c]{@{}l@{}}7.3\\ 10.1\\ 10\end{tabular} & \begin{tabular}[c]{@{}l@{}}13.6\\ 19.7\\ 19\end{tabular} & \begin{tabular}[c]{@{}l@{}}11.9\\ 12.5\\ 5.3\end{tabular} \\ \hline
IT & \begin{tabular}[c]{@{}l@{}}4.8\\ 5.7\\ 5.7\end{tabular} & \begin{tabular}[c]{@{}l@{}}2.4\\ 2.1\\ 1.6\end{tabular} & \begin{tabular}[c]{@{}l@{}}10.2\\ 10.4\\ 10\end{tabular} & \begin{tabular}[c]{@{}l@{}}7.3\\ 7.3\\ 8.5\end{tabular} & & \begin{tabular}[c]{@{}l@{}}5.4\\ 5.8\\ 8.8\end{tabular} & \begin{tabular}[c]{@{}l@{}}2.6\\ 3.7\\ 6.7\end{tabular} & \begin{tabular}[c]{@{}l@{}}4.8\\ 4.4\\ 5\end{tabular} & \begin{tabular}[c]{@{}l@{}}6.4\\ 10.1\\ 8.9\end{tabular} & \begin{tabular}[c]{@{}l@{}}10.1\\ 6.7\\ 8.5\end{tabular} & \begin{tabular}[c]{@{}l@{}}2.7\\ 4.8\\ 3.5\end{tabular} & \begin{tabular}[c]{@{}l@{}}1.6\\ 1.5\\ 2.8\end{tabular} & \begin{tabular}[c]{@{}l@{}}6\\ 5.4\\ 6.9\end{tabular} & \begin{tabular}[c]{@{}l@{}}3.2\\ 4\\ 3.9\end{tabular} & \begin{tabular}[c]{@{}l@{}}5.7\\ 5.7\\ 7.2\end{tabular} & \begin{tabular}[c]{@{}l@{}}4.1\\ 2.9\\ 3\end{tabular} \\ \hline
UK & \begin{tabular}[c]{@{}l@{}}8.2\\ 7.1\\ 8.2\end{tabular} & \begin{tabular}[c]{@{}l@{}}6.1\\ 5.7\\ 6.1\end{tabular} & \begin{tabular}[c]{@{}l@{}}5.8\\ 6.4\\ 7.3\end{tabular} & \begin{tabular}[c]{@{}l@{}}8.5\\ 9.1\\ 9.3\end{tabular} & \begin{tabular}[c]{@{}l@{}}7.2\\ 7.3\\ 10.1\end{tabular} & & \begin{tabular}[c]{@{}l@{}}5.5\\ 6.6\\ 7.6\end{tabular} & \begin{tabular}[c]{@{}l@{}}3.7\\ 7.3\\ 6.1\end{tabular} & \begin{tabular}[c]{@{}l@{}}5.5\\ 6.2\\ 6.7\end{tabular} & \begin{tabular}[c]{@{}l@{}}10.7\\ 8.4\\ 8.3\end{tabular} & \begin{tabular}[c]{@{}l@{}}15\\ 11.8\\ 12.8\end{tabular} & \begin{tabular}[c]{@{}l@{}}5.5\\ 4.5\\ 3.7\end{tabular} & \begin{tabular}[c]{@{}l@{}}4.7\\ 6\\ 6.8\end{tabular} & \begin{tabular}[c]{@{}l@{}}4.4\\ 6.4\\ 7.4\end{tabular} & \begin{tabular}[c]{@{}l@{}}5.7\\ 11.2\\ 10.2\end{tabular} & \begin{tabular}[c]{@{}l@{}}7.8\\ 4.5\\ 4.1\end{tabular} \\ \hline
CA & \begin{tabular}[c]{@{}l@{}}10.4\\ 9.3\\ 8\end{tabular} & \begin{tabular}[c]{@{}l@{}}12.4\\ 8.1\\ 8.3\end{tabular} & \begin{tabular}[c]{@{}l@{}}6.1\\ 5.7\\ 5.6\end{tabular} & \begin{tabular}[c]{@{}l@{}}4.3\\ 3.6\\ 3.7\end{tabular} & \begin{tabular}[c]{@{}l@{}}3.2\\ 3.8\\ 5.7\end{tabular} & \begin{tabular}[c]{@{}l@{}}5.2\\ 5.5\\ 5.6\end{tabular} & & \begin{tabular}[c]{@{}l@{}}6.1\\ 5.5\\ 4.4\end{tabular} & \begin{tabular}[c]{@{}l@{}}4.6\\ 4.3\\ 3.2\end{tabular} & \begin{tabular}[c]{@{}l@{}}5.7\\ 3.8\\ 2.6\end{tabular} & \begin{tabular}[c]{@{}l@{}}7.8\\ 8.2\\ 5.8\end{tabular} & \begin{tabular}[c]{@{}l@{}}6.2\\ 4.5\\ 3.6\end{tabular} & \begin{tabular}[c]{@{}l@{}}3.3\\ 3.9\\ 5\end{tabular} & \begin{tabular}[c]{@{}l@{}}7.3\\ 9.9\\ 6\end{tabular} & \begin{tabular}[c]{@{}l@{}}7.1\\ 3.3\\ 4.8\end{tabular} & \begin{tabular}[c]{@{}l@{}}7.8\\ 5.6\\ 5.3\end{tabular} \\ \hline
JP & \begin{tabular}[c]{@{}l@{}}3.3\\ 2.9\\ 3.2\end{tabular} & \begin{tabular}[c]{@{}l@{}}5.2\\ 6.5\\ 4.3\end{tabular} & \begin{tabular}[c]{@{}l@{}}2\\ 2.4\\ 3.1\end{tabular} & \begin{tabular}[c]{@{}l@{}}4.6\\ 3.2\\ 3.1\end{tabular} & \begin{tabular}[c]{@{}l@{}}2.6\\ 2.7\\ 2.5\end{tabular} & \begin{tabular}[c]{@{}l@{}}1.5\\ 3.5\\ 2.6\end{tabular} & \begin{tabular}[c]{@{}l@{}}2.6\\ 3.2\\ 2.5\end{tabular} & & \begin{tabular}[c]{@{}l@{}}1.6\\ 1.9\\ 1.9\end{tabular} & \begin{tabular}[c]{@{}l@{}}2\\ 4.1\\ 2.6\end{tabular} & \begin{tabular}[c]{@{}l@{}}2.5\\ 2\\ 2.7\end{tabular} & \begin{tabular}[c]{@{}l@{}}11.1\\ 9.8\\ 8.8\end{tabular} & \begin{tabular}[c]{@{}l@{}}5\\ 4.4\\ 4.2\end{tabular} & \begin{tabular}[c]{@{}l@{}}2.2\\ 0.5\\ 1\end{tabular} & \begin{tabular}[c]{@{}l@{}}3.1\\ 3.3\\ 1.4\end{tabular} & \begin{tabular}[c]{@{}l@{}}4.1\\ 4.3\\ 2.1\end{tabular} \\ \hline
ES & \begin{tabular}[c]{@{}l@{}}4.5\\ 3.9\\ 3.6\end{tabular} & \begin{tabular}[c]{@{}l@{}}1.3\\ 1.6\\ 1.8\end{tabular} & \begin{tabular}[c]{@{}l@{}}6.1\\ 6.6\\ 6.2\end{tabular} & \begin{tabular}[c]{@{}l@{}}3\\ 4.7\\ 4.3\end{tabular} & \begin{tabular}[c]{@{}l@{}}5.1\\ 9.2\\ 7.2\end{tabular} & \begin{tabular}[c]{@{}l@{}}3.3\\ 4.5\\ 4.7\end{tabular} & \begin{tabular}[c]{@{}l@{}}2.9\\ 3.7\\ 3\end{tabular} & \begin{tabular}[c]{@{}l@{}}2.4\\ 2.8\\ 3\end{tabular} & & \begin{tabular}[c]{@{}l@{}}4.4\\ 3.2\\ 4\end{tabular} & \begin{tabular}[c]{@{}l@{}}3.4\\ 3.5\\ 3.4\end{tabular} & \begin{tabular}[c]{@{}l@{}}2\\ 2.1\\ 3.5\end{tabular} & \begin{tabular}[c]{@{}l@{}}7\\ 7.1\\ 6\end{tabular} & \begin{tabular}[c]{@{}l@{}}2.4\\ 3.5\\ 1.9\end{tabular} & \begin{tabular}[c]{@{}l@{}}3.7\\ 5.5\\ 6.1\end{tabular} & \begin{tabular}[c]{@{}l@{}}2.9\\ 4.7\\ 3.8\end{tabular} \\ \hline
RU & \begin{tabular}[c]{@{}l@{}}3\\ 2.6\\ 2.6\end{tabular} & \begin{tabular}[c]{@{}l@{}}0.2\\ 1\\ 1\end{tabular} & \begin{tabular}[c]{@{}l@{}}5.4\\ 4.2\\ 3.6\end{tabular} & \begin{tabular}[c]{@{}l@{}}6\\ 4.1\\ 4.1\end{tabular} & \begin{tabular}[c]{@{}l@{}}6.6\\ 3.5\\ 4.1\end{tabular} & \begin{tabular}[c]{@{}l@{}}5.2\\ 3.5\\ 3.5\end{tabular} & \begin{tabular}[c]{@{}l@{}}2.9\\ 1.9\\ 1.5\end{tabular} & \begin{tabular}[c]{@{}l@{}}2.4\\ 3.5\\ 2.5\end{tabular} & \begin{tabular}[c]{@{}l@{}}3.6\\ 1.9\\ 2.4\end{tabular} & & \begin{tabular}[c]{@{}l@{}}2.8\\ 2\\ 2.4\end{tabular} & \begin{tabular}[c]{@{}l@{}}1.3\\ 0.6\\ 2.6\end{tabular} & \begin{tabular}[c]{@{}l@{}}3.3\\ 5\\ 3\end{tabular} & \begin{tabular}[c]{@{}l@{}}5.1\\ 3.3\\ 4.8\end{tabular} & \begin{tabular}[c]{@{}l@{}}4.2\\ 2.6\\ 2.2\end{tabular} & \begin{tabular}[c]{@{}l@{}}1.2\\ 0.4\\ 1\end{tabular} \\ \hline
AU & \begin{tabular}[c]{@{}l@{}}3.8\\ 2.7\\ 3.1\end{tabular} & \begin{tabular}[c]{@{}l@{}}6.9\\ 7.1\\ 7.8\end{tabular} & \begin{tabular}[c]{@{}l@{}}1.2\\ 2.4\\ 1.8\end{tabular} & \begin{tabular}[c]{@{}l@{}}2.4\\ 3\\ 2.4\end{tabular} & \begin{tabular}[c]{@{}l@{}}1.7\\ 2.6\\ 2\end{tabular} & \begin{tabular}[c]{@{}l@{}}7.1\\ 5.1\\ 6.3\end{tabular} & \begin{tabular}[c]{@{}l@{}}3.9\\ 4.2\\ 3.8\end{tabular} & \begin{tabular}[c]{@{}l@{}}2.8\\ 1.8\\ 3\end{tabular} & \begin{tabular}[c]{@{}l@{}}2.7\\ 2.1\\ 2.4\end{tabular} & \begin{tabular}[c]{@{}l@{}}2.8\\ 2.1\\ 2.8\end{tabular} & & \begin{tabular}[c]{@{}l@{}}2\\ 1.5\\ 2.1\end{tabular} & \begin{tabular}[c]{@{}l@{}}1.7\\ 1.7\\ 2\end{tabular} & \begin{tabular}[c]{@{}l@{}}2.2\\ 1.9\\ 1\end{tabular} & \begin{tabular}[c]{@{}l@{}}2.3\\ 1.4\\ 3.7\end{tabular} & \begin{tabular}[c]{@{}l@{}}1.6\\ 1.6\\ 2.7\end{tabular} \\ \hline
KR & \begin{tabular}[c]{@{}l@{}}3.8\\ 3.9\\ 3.9\end{tabular} & \begin{tabular}[c]{@{}l@{}}4.1\\ 5\\ 4.1\end{tabular} & \begin{tabular}[c]{@{}l@{}}0.8\\ 1.2\\ 1.1\end{tabular} & \begin{tabular}[c]{@{}l@{}}0.9\\ 1\\ 1.7\end{tabular} & \begin{tabular}[c]{@{}l@{}}0.6\\ 0.6\\ 1.1\end{tabular} & \begin{tabular}[c]{@{}l@{}}1.5\\ 1.4\\ 1.3\end{tabular} & \begin{tabular}[c]{@{}l@{}}1.8\\ 1.7\\ 1.7\end{tabular} & \begin{tabular}[c]{@{}l@{}}7.4\\ 6.3\\ 7.1\end{tabular} & \begin{tabular}[c]{@{}l@{}}0.9\\ 0.9\\ 1.7\end{tabular} & \begin{tabular}[c]{@{}l@{}}0.7\\ 0.4\\ 2.2\end{tabular} & \begin{tabular}[c]{@{}l@{}}1.1\\ 1.1\\ 1.5\end{tabular} & & \begin{tabular}[c]{@{}l@{}}0\\ 1\\ 0.9\end{tabular} & \begin{tabular}[c]{@{}l@{}}0.5\\ 0.3\\ 2.1\end{tabular} & \begin{tabular}[c]{@{}l@{}}0.8\\ 0.5\\ 0.5\end{tabular} & \begin{tabular}[c]{@{}l@{}}5.3\\ 5.4\\ 6.2\end{tabular} \\ \hline
PL & \begin{tabular}[c]{@{}l@{}}2.3\\ 1.9\\ 1.7\end{tabular} & \begin{tabular}[c]{@{}l@{}}0.6\\ 0.5\\ 1.1\end{tabular} & \begin{tabular}[c]{@{}l@{}}2.7\\ 2.1\\ 2.8\end{tabular} & \begin{tabular}[c]{@{}l@{}}3.1\\ 3.2\\ 3.6\end{tabular} & \begin{tabular}[c]{@{}l@{}}2.1\\ 1.9\\ 2.7\end{tabular} & \begin{tabular}[c]{@{}l@{}}1.3\\ 1.7\\ 2.3\end{tabular} & \begin{tabular}[c]{@{}l@{}}0.9\\ 1.3\\ 2.3\end{tabular} & \begin{tabular}[c]{@{}l@{}}3.3\\ 2.6\\ 3.3\end{tabular} & \begin{tabular}[c]{@{}l@{}}3.1\\ 2.7\\ 2.9\end{tabular} & \begin{tabular}[c]{@{}l@{}}1.8\\ 3.4\\ 2.5\end{tabular} & \begin{tabular}[c]{@{}l@{}}0.9\\ 1.1\\ 1.4\end{tabular} & \begin{tabular}[c]{@{}l@{}}0\\ 0.9\\ 0.9\end{tabular} & & \begin{tabular}[c]{@{}l@{}}1\\ 1.7\\ 2.7\end{tabular} & \begin{tabular}[c]{@{}l@{}}1.1\\ 1.9\\ 1.8\end{tabular} & \begin{tabular}[c]{@{}l@{}}0\\ 0.9\\ 2.9\end{tabular} \\ \hline
IL & \begin{tabular}[c]{@{}l@{}}6\\ 5\\ 4.5\end{tabular} & \begin{tabular}[c]{@{}l@{}}0.9\\ 0.5\\ 1\end{tabular} & \begin{tabular}[c]{@{}l@{}}2.4\\ 2.5\\ 1.8\end{tabular} & \begin{tabular}[c]{@{}l@{}}2.3\\ 2.7\\ 2.3\end{tabular} & \begin{tabular}[c]{@{}l@{}}1.6\\ 1.7\\ 1.2\end{tabular} & \begin{tabular}[c]{@{}l@{}}1.6\\ 2.2\\ 2.1\end{tabular} & \begin{tabular}[c]{@{}l@{}}2.8\\ 4\\ 2.2\end{tabular} & \begin{tabular}[c]{@{}l@{}}2\\ 0.4\\ 0.6\end{tabular} & \begin{tabular}[c]{@{}l@{}}1.5\\ 1.6\\ 0.8\end{tabular} & \begin{tabular}[c]{@{}l@{}}3.9\\ 2.7\\ 3.2\end{tabular} & \begin{tabular}[c]{@{}l@{}}1.7\\ 1.5\\ 0.6\end{tabular} & \begin{tabular}[c]{@{}l@{}}0.7\\ 0.4\\ 1.7\end{tabular} & \begin{tabular}[c]{@{}l@{}}1.3\\ 2.1\\ 2.2\end{tabular} & & \begin{tabular}[c]{@{}l@{}}4.2\\ 3.6\\ 1.6\end{tabular} & \begin{tabular}[c]{@{}l@{}}1.6\\ 0.7\\ 1.7\end{tabular} \\ \hline
NL & \begin{tabular}[c]{@{}l@{}}2.7\\ 2.6\\ 2.3\end{tabular} & \begin{tabular}[c]{@{}l@{}}0.5\\ 0.9\\ 0.9\end{tabular} & \begin{tabular}[c]{@{}l@{}}2\\ 1.8\\ 2.1\end{tabular} & \begin{tabular}[c]{@{}l@{}}3.7\\ 5.2\\ 4.3\end{tabular} & \begin{tabular}[c]{@{}l@{}}2.4\\ 2.4\\ 2.3\end{tabular} & \begin{tabular}[c]{@{}l@{}}1.8\\ 3.8\\ 2.9\end{tabular} & \begin{tabular}[c]{@{}l@{}}2.4\\ 1.3\\ 1.8\end{tabular} & \begin{tabular}[c]{@{}l@{}}2.4\\ 2.3\\ 0.9\end{tabular} & \begin{tabular}[c]{@{}l@{}}1.9\\ 2.6\\ 2.4\end{tabular} & \begin{tabular}[c]{@{}l@{}}2.8\\ 2.1\\ 1.5\end{tabular} & \begin{tabular}[c]{@{}l@{}}1.5\\ 1.1\\ 2.1\end{tabular} & \begin{tabular}[c]{@{}l@{}}1\\ 0.6\\ 0.4\end{tabular} & \begin{tabular}[c]{@{}l@{}}1.3\\ 2.3\\ 1.4\end{tabular} & \begin{tabular}[c]{@{}l@{}}3.6\\ 3.7\\ 1.6\end{tabular} & & \begin{tabular}[c]{@{}l@{}}1.2\\ 1.6\\ 0.7\end{tabular} \\ \hline
IN & \begin{tabular}[c]{@{}l@{}}2.3\\ 1.9\\ 2.7\end{tabular} & \begin{tabular}[c]{@{}l@{}}0.9\\ 1.3\\ 1.8\end{tabular} & \begin{tabular}[c]{@{}l@{}}1.2\\ 1\\ 1.4\end{tabular} & \begin{tabular}[c]{@{}l@{}}2.2\\ 2.6\\ 1.6\end{tabular} & \begin{tabular}[c]{@{}l@{}}1.2\\ 1\\ 1.3\end{tabular} & \begin{tabular}[c]{@{}l@{}}1.7\\ 1.2\\ 1.5\end{tabular} & \begin{tabular}[c]{@{}l@{}}1.8\\ 1.8\\ 2.7\end{tabular} & \begin{tabular}[c]{@{}l@{}}2.2\\ 2.3\\ 1.8\end{tabular} & \begin{tabular}[c]{@{}l@{}}1\\ 1.7\\ 2\end{tabular} & \begin{tabular}[c]{@{}l@{}}0.6\\ 0.3\\ 0.9\end{tabular} & \begin{tabular}[c]{@{}l@{}}0.8\\ 1\\ 2\end{tabular} & \begin{tabular}[c]{@{}l@{}}4.2\\ 4.5\\ 6.6\end{tabular} & \begin{tabular}[c]{@{}l@{}}0\\ 0.8\\ 3.1\end{tabular} & \begin{tabular}[c]{@{}l@{}}1\\ 0.5\\ 2.3\end{tabular} & \begin{tabular}[c]{@{}l@{}}0.8\\ 1.2\\ 1\end{tabular} & \\ \hline
\end{tabular}}
\caption{Bilateral collaborations: For a country in $x$-coordinate, the numbers correspond to the \% of its international collaborations with a country of the $y$-coordinate, for 2005, 2010, and 2015 }
\label{t_collaboration_2_countries}
\end{table}

In Table \ref{t_collaboration_2_countries}, we provide more specific information about bi-national collaboration.
More precisely, since our dataset is large enough, collaboration between two countries can be extracted. It corresponds to publications involving at least one author in a country $X$, and one author in a country $Y$. Additional authors and/or countries can also be involved. This number can then be divided either by the total number of international collaborations of the country $X$, or by the total number of international collaborations of the country $Y$. In the first case, it gives for the country $X$ the relative importance of collaborations with the country $Y$, while in the second case it gives for the country $Y$ the relative importance of collaborations with the country $X$. Table \ref{t_collaboration_2_countries} contains this information for 16 countries and for the three years. These three information are provided in each cell,
with 2005 on the top, 2010 in the middle, and 2015 on the bottom.

A general trend is visible in this table: For all countries except for China, the ratios of collaborations with the USA slightly decrease. On the other hand, for all countries, USA included, the ratios of collaborations with China increase, even if for most of them, collaborations with the USA are still in much higher numbers compared to the collaborations with China. In that respect, Australia is quite an exception, with a higher percentage of collaborations with China than with the USA. Note that for the USA, the ratio of collaborations with China has doubled during a period of 10 years (from 11.3\% to 22.2 \%). For other bi-national collaborations, involving a few \% of all international collaborations for both countries, a general trend is not clearly visible, and fluctuations are more important (also because the numbers of publications involved are smaller).

\section{Conclusion}

One of the unexpected outcomes of these investigations is the rapid increase of
the number of publications, but also of the significant change of
many predictors over a period of 10 years only. Indeed, if we
gather some results obtained in \eqref{eq_step_2}, \eqref{eq_authors},
\eqref{eq_institutes}, \eqref{eq_nationalities}, \eqref{eq_references_no_0},
\eqref{eq_pages}, \eqref{eq_keywords_no_0},
and in Table \ref{t_open}, we obtain Table \ref{t_increase}.
If we summarize in one sentence the content of this table, it would be that
the publications in mathematics are becoming more collaborative, more international, longer, with more references and keywords, more freely accessible, and
especially more numerous.

\begin{table}[h]
\centering
\begin{tabular}{|l|c|c|c|}
\hline
 & 2005 & 2010 & 2015 \\ \hline
\# publications & 45'035 & 62'945 & 76'788 \\ \hline
\# authors & 2.10 & 2.23 & 2.39 \\ \hline
\# institutes & 1.57 & 1.74 & 1.90 \\ \hline
\# countries & 1.23 & 1.28 & 1.32 \\ \hline
\# references & 19 & 21.7 & 24.5 \\ \hline
\# pages & 15.5 & 15.5 & 16.9 \\ \hline
\# keywords & 4.38 & 4.48 & 4.58 \\ \hline
open access (\%) & 16.4 & 21.2 & 32.3 \\ \hline
\end{tabular}
\caption{The increase of many predictors}
\label{t_increase}
\end{table}

For the citations, it is quite natural that the Journal Impact Factor plays an important role.
In that sense, its appearance as the most important predictor (whenever available) is not
surprising. For more general publications, the importance of the number of references is also not so surprising: quite often, a paper with numerous references corresponds to a paper which is well nested in the research landscape, and as a consequence it can be cited
by several authors.
The importance of other predictors is less clear, and no conclusion can be established for them. At this level, the real content of a paper certainly matters more than any bibliometric predictor.

The content of Section \ref{sec_countries} seems new. From the point of view of the authors, these statistics are maybe not so relevant, since it is difficult to move to another country, just to get more citations. On the other hand, at the level of the scientific policy of countries, knowing the relative ranking of countries with respect to several macro statistics is certainly important. In this direction, our results are only partial and preliminary, but we hope that it can trigger additional investigations in the future.

\section*{Acknowledgement}
The authors would like to thank the referees for their careful reading of a previous version 
of this work and for their numerous advises.
They also thank Keiko Koide, from the Technical Support / Customer Service of 
Clarivate (Asia Pacific), for having answered all their inquiries related to WoS database.

%
\bibliographystyle{apalike}
%

\begin{thebibliography}{}

\bibitem{AB}
Amodio, P. and Brugnano, L. (2014).
\newblock Recent advances in bibliometric indexes and the
PaperRank problem.
\newblock \emph{Journal of Computational and Applied Mathematics} 267, 182--194.

\bibitem{AC}
Aria, M. and Cuccurullob, C. (2017).
\newblock \emph{bibliometrix}: An R-tool for comprehensive science mapping
analysis.
\newblock \emph{Journal of Informetrics} 11, 959--975.

\bibitem{Bens}
Bensman, S., Smolinsky, L., and Pudovkin, A. (2010).
\newblock Mean Citation Rate per Article in Mathematics Journals:
Differences From the Scientific Model.
\newblock \emph{Journal of the American Society for information science and technology} 61(7), 1440--1463.

\bibitem{BL}
Behrens, H. and Luksch, P. (2011).
\newblock Mathematics 1868–2008: a bibliometric analysis.
\newblock \emph{Scientometrics} 86, 179--194.

\bibitem{BFOS}
Breiman, L., Friedman. J.H., Olshen, R.A., and Stone, C.J. (1984).
\newblock Classification and Regression Trees.
\newblock CHAPMAN \& HALL/CRC.

\bibitem{deB}
De Battisti, F. and Salini, S. (2013).
\newblock Robust analysis of bibliometric data.
\newblock \emph{Stat. Methods Appl.} 22, 269--283.

\bibitem{DT}
Didegah, F. and Thelwall, M. (2013).
\newblock Which factors help authors produce the highest impact research?
Collaboration, journal and document properties.
\newblock \emph{Journal of Informetrics} 7, 861--873.

\bibitem{Dunne}
Dunne, E. (2021).
\newblock Don't count on it.
\newblock \emph{Notices Amer. Math. Soc.} 68 no. 1, 114--118.

\bibitem{GDD}
Goldfinch, S., Dale, T., and DeRouen, K. (2003).
\newblock Science from the periphery: Collaboration, networks and `Periphery Effects' in the citation of New Zealand Crown
Research Institutes articles, 1995--2000.
\newblock \emph{Scientometrics} 57(3), 321--337.

\bibitem{G1}
Grossman, J.W. (2002).
\newblock The evolution of the mathematical research collaboration graph.
\newblock in Proceedings of the Thirty-third Southeastern International Conference on Combinatorics, Graph Theory and Computing (Boca Raton, FL, 2002).
\newblock \emph{Congr. Numer.} 158, 201--212.

\bibitem{G2}
Grossman, J.W. (2005).
\newblock Patterns of research in mathematics.
\newblock \emph{Notices Amer. Math. Soc.} 52 no. 1, 35--41.

\bibitem{LFSEM}
Luo, J., Flynn, J.M., Solnick, R.E., Ecklund, E.H., and Matthews, K.R.W. (2011).
\newblock International Stem Cell Collaboration: How Disparate Policies between the United States and the United Kingdom Impact Research.
\newblock \emph{PLoS ONE} 6(3), e17684.

\bibitem{OZ}
\"Ozkaya, A. (2018).
\newblock Bibliometric analysis of the studies in the field of
mathematics education.
\newblock \emph{Educational Research and Reviews} 13(22), 723--734.

\bibitem{PI}
Paik, J. and Rivin, I. (2020).
\newblock Bibliometric Analysis of Senior US Mathematics Faculty.
\newblock Preprint arXiv:2008.11196.

\bibitem{PVG}
Pedregosa, F., Varoquaux, G., Gramfort, A., Michel, V., Thirion, B., Grisel, O., Blondel, M., Prettenhofer, P., Weiss, R., Dubourg, V., Vanderplas, J., Passos, A., Cournapeau, D., Brucher, M., Perrot, M., and Duchesnay, E. (2011).
\newblock Scikit-learn: Machine Learning in Python.
\newblock \emph{Journal of Machine Learning Research} 12, 2825--2830.

\bibitem{RD}
Rusin, D. (2015).
\newblock A Gentle Introduction to the Mathematics Subject Classification Scheme.
\newblock Link provided by \\
https://en.wikipedia.org/wiki/Mathematics\_Subject\_Classification.

\bibitem{Gini}
Scikit-learn.org.
\newblock https://scikit-learn.org/stable/modules/tree.html

\bibitem{SWBA}
Smith, M.J., Weinberger, C., Bruna, E.M., and Allesina, S. (2014).
\newblock The Scientific Impact of Nations: Journal Placement and Citation Performance.
\newblock \emph{PLoS ONE} 9(10), e109195.

\bibitem{Szo}
Szomszor, M., Pendlebury, D.A., and Adams, J. (2020).
\newblock How much is too much? The difference between research
influence and self‑citation excess.
\newblock \emph{Scientometrics} 123, 1119--1147.

\bibitem{S1}
Sooryamoorthy, R. (2009).
\newblock Do types of collaboration change citation? Collaboration and citation patterns
of South African science publications.
\newblock \emph{Scientometrics} 81(1), 177--193.

\bibitem{S2}
Sooryamoorthy, R. (2017).
\newblock Do types of collaboration change citation? A scientometric analysis of social science publications in South Africa.
\newblock \emph{Scientometrics} 111, 379--400.

\bibitem{Verma}
Verma, R., Lobos-Ossand\'on, V., Merig\'o, J.M., Cancino, C., and Sienz, J. (2021).
\newblock Forty years of applied mathematical modelling: A bibliometric study.
\newblock \emph{Applied Mathematical Modelling} 89, 1177--1197.

\bibitem{WWL}
Wagner, C., Whetsell, T., and Leydesdorff, L. (2017).
\newblock Growth of international collaboration in science:
revisiting six specialties.
\newblock \emph{Scientometrics} 110, 1633--1652.

\bibitem{WTG}
Wang, L., Thijs, B., and Gl\"anzel, W. (2015).
\newblock Characteristics of international collaboration in sport sciences publications and its influence on citation impact.
\newblock \newblock{Scientometrics} 105, 843--862.

\bibitem{WZJZ}
Wang, M., Zhang, J., Jiao, S., and Zhang, T. (2019).
\newblock Evaluating the impact of citations of articles based on knowledge flow patterns hidden in the citations.
\newblock \emph{PLoS ONE} 14(11), e0225276.

\end{thebibliography}

\end{document}